# Efficient Graph Edit Distance Computation and Verification via Anchor-aware Lower Bound Estimation

Lijun Chang
The University of Sydney
Lijun.Chang@sydney.edu.au

Xing Feng
University of New South Wales
xingfeng@cse.unsw.edu.au

Xuemin Lin
University of New South Wales
lxue@cse.unsw.edu.au

Lu Qin
University of Technology Sydney
Lu.Qin@uts.edu.au

Wenjie Zhang
University of New South Wales
zhangw@cse.unsw.edu.au

October 1, 2017

## Abstract

Graph edit distance (GED) is an important similarity measure adopted in a similarity-based analysis between two graphs, and computing GED is a primitive operator in graph database analysis. Partially due to the NP-hardness, the existing algorithms for computing GED are only able to process very small graphs with less than 30 vertices. Motivated by this, in this paper we systematically investigate the problems of GED computation, and GED verification (*i.e.*, verify whether the GED between two graphs is no larger than a user-given threshold). Firstly, we develop a unified framework that can be instantiated into either a best-first search approach AStar$^+$ or a depth-first search approach DFS$^+$. Secondly, we design anchor-aware lower bound estimation techniques to compute tighter lower bounds for intermediate search states, which significantly reduce the search spaces of both AStar$^+$ and DFS$^+$. We also propose efficient algorithms to compute the lower bounds. Thirdly, based on our unified framework, we contrast AStar$^+$ with DFS$^+$ regarding their time and space complexities, and recommend that AStar$^+$ is better than DFS$^+$ by having a much smaller search space. Extensive empirical studies validate that AStar$^+$ performs better than DFS$^+$, and show that our AStar$^+$-BMa approach outperforms the state-of-the-art technique by more than four orders of magnitude.

## 1 Introduction

Graph model is becoming ubiquitous and has been used to model the complex relationships among entities in a wide spectrum of applications. Recent decades have witnessed significant research efforts towards many fundamental problems in managing and analyzing graph data. Among them, computing similarities between graphs is a fundamental and essential operation in many applications. For example, it serves as a key building block in graph classification [20], graph clustering [25], and similarity search and joins of graph databases [29, 30]. Moreover, computing the similarity between graphs also assists to identify functionally related en-zyme clusters in biochemistry [21], compare electroencephalogram in medicine [3], and retrieve similar objects in videos [7].

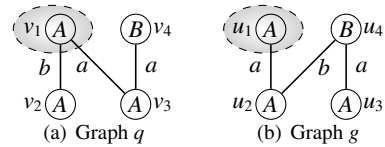

Figure 1: An example

Besides various similarity measures such as the maximum common subgraph [6, 10] and the number of miss-matching edges [32], the *graph edit distance* (GED) [11, 26, 29] has also been shown an important similarity measure since it gives the minimum amount of distortion needed to transform one graph into the other and it is a metric. As illustrated in [13], the GED between graphs $q$ and $g$, denoted by $\delta(q, g)$, equals the minimum *editorial cost* among all vertex mappings from $q$ to $g$, where the editorial cost of a mapping is the minimum number of *edit operations* involved to transform $q$ to $g$ by obeying the mapping. Here, an edit operation is a vertex insertion/deletion/relabeling, or edge insertion/deletion/relabeling. For example, regarding $q$ in Figure 1(a) and $g$ in Figure 1(b), it can be verified that the mapping $\{v_1 \mapsto u_1, v_2 \mapsto u_2, v_3 \mapsto u_3, v_4 \mapsto u_4\}$ has an editorial cost of 3, and it has the minimum editorial cost among all mappings. Thus, the GED between $q$ and $g$ is $\delta(q, g) = 3$.

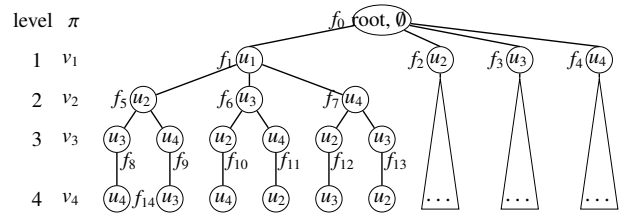

Figure 2: Search tree $\mathcal{T}$

**Existing Approaches to GED Computation.** As computing GED is NP-hard [28], the existing approaches conduct searches on *a search tree*, which represents all vertex mapping from $q$ to $g$ in a prefix-shared manner according to a matching order of vertices $V(q)$ of $q$, for finding the mapping with the minimum editorial cost. Given the matching order $\pi = (v_1, v_2, v_3, v_4)$ of $q$, the search tree $\mathcal{T}$ for computing $\delta(q, g)$ in Figure 1 is shown in Figure 2. Each node at level $i$ of $\mathcal{T}$ represents a (partial) mapping from $(v_1, \ldots, v_i)$ to vertices of $g$ (i.e., represents an intermediate search state). All the full mappings are represented by the leaf nodes of $\mathcal{T}$. For example, node $f_{14}$ represents the mapping $\{v_1 \mapsto u_1, v_2 \mapsto u_2, v_3 \mapsto u_3, v_4 \mapsto u_4\}$. Based on the parent-child relationships of nodes in a rooted tree, parent-child (as well as ancestor-descendant) relationships are also defined for mappings in $\mathcal{T}$. To avoid an exhaustive search of all mappings in $\mathcal{T}$, *a lower bound cost* is computed for each partial mapping $f$, denoted by $\delta(f)$, which is a lower bound of the editorial cost of all full mappings that are descendants of $f$ in $\mathcal{T}$.

Table 1: Our GED Computation Algorithms (only the lower bound $\delta^{\text{LS}}(\cdot)$ has been used in existing works)

| Search Paradigms | Lower Bounds of Partial Mappings | | | | | |
|---|---|---|---|---|---|---|
| | $\delta^{\text{BMa}}(\cdot)$ | $\delta^{\text{BM}}(\cdot)$ | $\delta^{\text{BMaN}}(\cdot)$ | $\delta^{\text{SMa}}(\cdot)$ | $\delta^{\text{LSa}}(\cdot)$ | $\delta^{\text{LS}}(\cdot)$ |
| AStar$^+$ | AStar$^+$-BMa | AStar$^+$-BM | AStar$^+$-BMaN | AStar$^+$-SMa | AStar$^+$-LSa | AStar$^+$-LS (improves A*GED [24]) |
| DFS$^+$ | DFS$^+$-BMa | DFS$^+$-BM | DFS$^+$-BMaN | DFS$^+$-SMa | DFS$^+$-LSa | DFS$^+$-LS (improves DF_GED [5]) |

The existing techniques conduct searches on the search tree in two fashions: best-first search and depth-first search.

*Best-First Search.* The existing best-first search approaches in [23, 24, 29, 30] are essentially the same, which refer to the algorithm A*GED in [24]. A*GED maintains a search frontier of $\mathcal{T}$ by a priority queue $Q$ that initially contains only the root (i.e., an empty mapping) of $\mathcal{T}$, and runs iteratively. In each iteration, A*GED pops from $Q$ the mapping that has the minimum lower bound cost, and pushes all its children into $Q$. It is guaranteed that the first full mapping popped from $Q$ has the minimum editorial cost among all full mappings from $q$ to $g$, and thus has editorial cost $\delta(q, g)$. For example, consider the search tree in Figure 2. Initially, the search frontier consists of $f_0$ (i.e., $Q = \{f_0\}$). In the first iteration, $f_0$ is popped from $Q$, and its children are pushed into $Q$; thus after the first iteration, we have $Q = \{f_1, f_2, f_3, f_4\}$. Assume $f_1$ has the minimum lower bound cost in $Q$. In the second iteration, $f_1$ is popped from $Q$, and its children are pushed into $Q$; thus $Q = \{f_5, f_6, f_7, f_2, f_3, f_4\}$. So on so forth, the algorithm terminates after popping the first full mapping (i.e., $f_{14}$) from $Q$.

A*GED uses a *label set-based lower bound*, denote by $\delta^{\text{LS}}(\cdot)$, for partial mappings. Given a partial mapping $f$, let $q \backslash f$ and $g \backslash f$ denote the unmapped parts of $q$ and $g$, respectively. Then, $\delta^{\text{LS}}(f)$ is defined as the sum of (1) the difference between the vertex labels of $q \backslash f$ and $g \backslash f$, and (2) the difference between the edge labels of $q \backslash f$ and $g \backslash f$. For example, given the partial mapping $f_1 = \{v_1 \mapsto u_1\}$ for the graphs in Figure 1, the multi-sets of vertex labels and edge labels of $q \backslash f_1$ are $\{A, A, B\}$ and $\{a, a, b\}$, respectively, which are the same as that of $g \backslash f_1$; thus, we have $\delta^{\text{LS}}(f_1) = 0$.

*Depth-First Search.* In view of the large memory consumptions of A*GED that keeps all intermediate search states in $Q$ in main memory, recent studies such as DF_GED [1, 5] and CSI_GED [11] suggest to conduct a depth-first search on the search tree $\mathcal{T}$. An upper bound $\bar{\delta}(q, g)$ of $\delta(q, g)$ is maintained to be the minimum editorial cost among all visited full mappings. DF_GED [1, 5] visits mappings in $\mathcal{T}$ in a depth-first manner, and uses the label set-based lower bound $\delta^{\text{LS}}(\cdot)$ for pruning. That is, when visiting a mapping $f$, if $\delta^{\text{LS}}(f) \geq \bar{\delta}(q, g)$, then the entire subtree of $\mathcal{T}$ rooted at $f$ is pruned. Otherwise, DF_GED visits the children of $f$ in non-decreasing order with respect to their lower bound costs. For example, consider the search tree in Figure 2 and assume that $\delta^{\text{LS}}(f_1) < \delta^{\text{LS}}(f_2) < \delta^{\text{LS}}(f_3) < \delta^{\text{LS}}(f_4)$. After visiting the root $f_0$ of $\mathcal{T}$, DF_GED first visits all mappings in the subtree rooted at $f_1$, then all mappings in the subtree rooted at $f_2$, and so forth. CSI_GED [11] is similar to but differs from DF_GED by enumerating edge mappings and using a degree-based lower bound.

**Our Contributions.** In this paper, we systematically study the problems of GED computation, and GED verification (i.e., verify whether the GED between two graphs is no larger than a user-given threshold $\tau$). A summary of our GED computation algorithms is shown in Table 1. Our main contributions are summarized as follows.

① *A Unified Framework* (Section 3). We develop a unified framework that can be instantiated into either a best-first search approach AStar$^+$ or a depth-first search approach DFS$^+$. Besides, any lower bound estimation technique for partial mappings can be adopted. Our paradigm of AStar$^+$ is similar to but differs from that of A*GED. AStar$^+$ aims to reduce the main memory consumption, as follows. Firstly, it stores each intermediate search state in a constant main memory space. Secondly, it utilizes an upper bound of $\delta(q, g)$ to prune the priority queue. On the other hand, our paradigm of DFS$^+$ is similar to that of DF_GED.

② *Anchor-aware Tighter Lower Bounds and Efficient Computation* (Section 4). We notice that the tightness as well as the computational efficiency of the lower bounds of partial mappings have a dramatic impact on the performance of a GED algorithm. Thus, we design anchor-aware techniques to compute tighter lower bounds, which significantly reduce the search spaces of both AStar$^+$ and DFS$^+$. Given a partial mapping $f$ from $q$ to $g$, we categorize the vertices of $q$ and $g$ into *anchored vertices* (i.e., in $f$) and *free vertices* (i.e., not in $f$).

Firstly, we propose an *anchor-aware label set-based lower bound*, denoted by $\delta^{\text{LSa}}(\cdot)$. It improves $\delta^{\text{LS}}(\cdot)$ by partitioning the edges of $q \backslash f$ and $g \backslash f$, respectively, into cross/adjacent edges of each anchored vertex, and other edges as inner edges. Then, the difference between the edge labels of $q \backslash f$ and $g \backslash f$, used in $\delta^{\text{LS}}(\cdot)$, is refined as the difference between the edge labels of each component of the partitions. Thus, $\delta^{\text{LSa}}(f) \geq \delta^{\text{LS}}(f)$ holds for any mapping $f$. For example, continuing the above example of computing $\delta^{\text{LS}}(f_1)$, any full mapping extending $f_1$ must edit the adjacent edges $\{(v_1, v_2), (v_1, v_3)\}$ of $v_1$ to match the adjacent edges $\{(u_1, u_2)\}$ of $u_1$ and edit the inner edges $\{(v_4, v_3)\}$ to match the inner edges $\{(u_2, u_4), (u_3, u_4)\}$; thus, the lower bound is computed as $\delta^{\text{LSa}}(f_1) = 2$, which is larger than $\delta^{\text{LS}}(f_1)$. More-over, we also propose efficient techniques to compute the lower bound costs of all children of a partial mapping, regarding $\delta^{\text{LS}}(\cdot)$ and $\delta^{\text{LSa}}(\cdot)$, in totally linear time with respect to the sizes of $q$ and $g$; note that, the existing techniques (*e.g.*, A*GED) take quadratic time.

Secondly, we for the first time adopt the branch-match based lower bound proposed in [31] to GED computation, and we further improve it by our anchor-aware technique, denoted by $\delta^{\text{BMa}}(\cdot)$. We prove that $\delta^{\text{BMa}}(f) \geq \delta^{\text{LSa}}(f)$ holds for any partial mapping $f$. We also propose techniques to compute the lower bound costs of all children of a partial mapping, regarding $\delta^{\text{BMa}}(\cdot)$, in $O((|V(q)| + |V(g)|)^3)$ total time.

③ *Contrast* AStar+ *with* DFS+ (Section 5). Based on our unified framework, we theoretically contrast AStar+ with DFS+, and show that AStar+ has a smaller search space than DFS+ when regarding the same lower bound estimation technique. Moreover, *our tight lower bound $\delta^{\text{BMa}}(\cdot)$ significantly reduces the search space as well as the memory consumption of AStar+*. Thus, we recommend that AStar+ is better than DFS+, which is more evident for GED computation. The gap between AStar+ and DFS+ for GED verification is not as significant as for GED computation. This is because the user-given threshold $\tau$ significantly reduces the search space of DFS+, and the search spaces of DFS+ and AStar+ are the same if the two graphs are dissimilar (*i.e.*, $\delta(q, g) > \tau$). Note that, (1) the existing work in [11] argues that depth-first search is better than best-first search and thus uses CSI_GED for both GED computation and GED verification, and (2) the existing works on graph similarity search in [17, 27, 29, 30, 31] all use A*GED for GED verification.

④ *Extensive Performance Studies* (Section 6). We conduct extensive performance studies on real graphs, which validate that AStar+ performs better than DFS+ and thus debunk the recent claims in [5, 11] that depth-first search is more suitable than best-first search for GED computation. The results show that our AStar+-BMa approach (*i.e.*, AStar+ with the lower bound $\delta^{\text{BMa}}(\cdot)$) has a very small search space, and outperforms the state-of-the-art techniques CSI_GED and DF_GED by more than four orders of magnitude. Moreover, it is interesting to observe that our AStar+-LS approach (*i.e.*, AStar+ with the lower bound $\delta^{\text{LS}}(\cdot)$), which can be considered as an improved version of A*GED by reducing memory consumption and speeding up lower bound computation, also performs better than CSI_GED; note that, CSI_GED is reported in [11] to outperform A*GED by over two orders of magnitude.

**Related Works.** Related works are categorized as below.

*(1) Compute* GED. The notion of graph edit distance (GED) was firstly proposed in [26] to quantify the distance between two graphs. Zeng *et al.* [28] proved that computing the exact GED between two graphs is NP-hard. Riesen *et al.* [24] designed a best-first search algorithm A*GED to compute GED. In view of the large memory consumption of A*GED, depth-first search algorithms DF_GED [1, 5] and CSI_GED [11] are recently proposed and shown to outperform the best-first search approach A*GED. However, all the existing algorithms cannot process graphs with more than 30 vertices. In this paper, we illustrate that best-first search is more suitable than depth-first search for GED computation. Moreover, we develop anchor-aware tighter lower bounds and also efficient computation techniques to significantly reduce the search space of our best-first search paradigm AStar+, such that our AStar+-BMa algorithm outperforms the state-of-the-art algorithm CSI_GED by more than four orders of magnitude.

*(2) Graph Similarity Search.* GED-based graph similarity search is studied in [17, 27, 29, 30, 31]; that is, given a graph database $D$ and a query graph $g$, find all graphs $g'$ in $D$ such that the GED between $g$ and $g'$ is no larger than a user-given threshold (*i.e.*, $g'$ is similar to $g$). All these works mainly focus on designing effective index structures — *e.g.*, q-gram-based index [30], star structure-based index [27], and subgraph-based index [17, 29] — to filter out from $D$ as many dissimilar graphs to $g$ as possible, while all remaining candidates are verified by A*GED. In this paper, we propose a new algorithm AStar+-BMa for GED verification that outperforms A*GED by several orders of magnitude. Thus, AStar+-BMa should be adopted for GED verification in future researches on graph similarity search. Moreover, existing indexing techniques on graph similarity search may need to be reevaluated, as a result of our significantly improved GED verification algorithm.

*(3) Compute Maximum Common Subgraph.* Measuring the similarity between two graphs based on their maximum common subgraph is also studied in the literature. As computing the maximum common subgraph is NP-hard, Mcgregor [18] proposed a depth-first search method, while more advanced pruning techniques are later proposed in [2, 15]. Another strategy is first constructing a product graph of the two input graphs, and then computing the maximum clique of the product graph which corresponds to the maximum common subgraph of these two graphs [14, 22]. These techniques cannot be applied to GED computation, since a vertex in one graph can map to any vertex in another graph for computing GED while a vertex in one graph can only map to vertices with the same label in another graph when computing maximum common subgraph.

## 2 Preliminaries

In this paper, we focus on labeled and undirected graphs $G = (V, E, l)$, where $V$ is the set of vertices, $E \subseteq V \times V$ is the set of edges, and $l : V \cup E \rightarrow \Sigma$ is a labelling function that assigns each vertex and/or edge in $G$ a label from $\Sigma$; that is, $l(u)$ and $l(u, u')$ are the labels of vertex $u$ and edge $(u, u')$, respectively. We denote the degree of vertex $u$ by $d(u)$. Given a graph $g$, we denote its vertex set and edge set by $V(g)$ and $E(g)$, respectively, and denote its number of vertices and number of edges by $|V(g)|$ and $|E(g)|$, respectively. We denote the size of $g$ by $\text{size}(g) = |V(g)| + |E(g)|$. Given a vertex subset $V_s \subseteq V(g)$, the subgraph of $g$ induced by $V_s$ is $g[V_s] = (V_s, \{(u, u') \in E(g) \mid u, u' \in V_s\}, l)$. In the following, for presentation simplicity we refer to a labeled and undirected graph simply as a graph.

*Definition 2.1:* A graph $q$ is **isomorphic** to another graph $g$ if there exists a *bijective* mapping $f$ from $V(q)$ to $V(g)$ such that (1) $l(v) = l(f(v))$ for each $v \in V(q)$, (2) $(v, v')$ is in $E(q)$ if and only if $(f(v), f(v'))$ is in $E(g)$, $\forall v, v' \in V(q)$, and (3) moreover

$l(v, v') = l(f(v), f(v'))$ for each $(v, v') \in E(q)$.

A *graph edit operation* on a graph is an operation that transforms the graph. Specifically, it includes the following six edit operations: inserting/deleting an isolated vertex into/from the graph (*vertex insertion* and *vertex deletion*), adding/deleting an edge between two vertices (*edge insertion* and *edge deletion*), and changing the label of a vertex/edge (*vertex relabeling* and *edge relabeling*). In this paper, we consider the case of uniform cost that the edit operations have the same cost.[1]

*Definition 2.2:* The **graph edit distance** (GED) between two graphs $q$ and $g$, denoted by $\delta(q, g)$, is the minimum number of edit operations that can transform $q$ to be isomorphic to $g$.

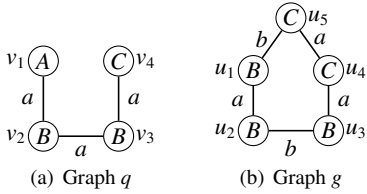

(a) Graph $q$      (b) Graph $g$
Figure 3: Sample graphs

Note that, GED is a metric [26] and $\delta(q, g) = \delta(g, q)$. Consider the two graphs $q$ and $g$ in Figure 3, where vertex labels are illustrated inside circles (*i.e.*, $A, B, C$) and edge labels are illustrated beside edges (*i.e.*, $a, b$). One possible sequence of edit operations for transforming $q$ to be isomorphic to $g$ is as follows, (1) change the label of vertex $v_1$ from $A$ to $B$, (2) change the label of edge $(v_2, v_3)$ from $a$ to $b$, (3) insert an isolated vertex $v_5$ with label $C$, (4) add an edge with label $b$ between $v_1$ and $v_5$, and (5) add an edge with label $a$ between $v_4$ and $v_5$. Thus, the GED between $q$ and $g$ is at most 5. Nevertheless, computing the exact GED is NP-hard [28].

**Problem Statement.** Given two graphs $q$ and $g$, we study

1. the problem of GED **computation** that computes the GED, $\delta(q, g)$, between $q$ and $g$, and

2. the problem of GED **verification** that verifies whether $\delta(q, g) \leq \tau$ for a user-given threshold $\tau$.

In the following, we present our techniques by mainly focusing on GED computation, while GED verification is discussed in Section 5.3. We use $v$ and its variants, $v', v_1, v_2, \ldots$, to denote vertices in $q$, and use $u$ and its variants, $u', u_1, u_2, \ldots$, to denote vertices in $g$. Frequently used notations are summarized in Table 2.

## 2.1 GED Computation via Vertex Mapping

We start with some simplifications, and then present the idea of GED computation via enumerating vertex mappings.

**Simplifications.** As GED is a metric, any of the two graphs can be regarded as $q$. In our algorithms, we choose $q$ to be the graph with fewer vertices (*i.e.*, $|V(q)| \leq |V(g)|$); if a tie occurs, an arbitrary one is chosen. We prove in the following lemma that, there is no vertex deletion in the optimal sequence of edit

---

[1]While the general idea presented in this paper also works for non-uniform costs, a detailed theoretical and experimental analysis of extending our techniques to the case of non-uniform costs will be a future work.

Table 2: Frequently used notations

| Notation | Description |
|---|---|
| $q, g$ | Two graphs |
| $V(g), E(g)$ | The vertex set and edge set of $g$ |
| $\delta(q, g)$ | The GED between graphs $q$ and $g$ |
| $\bar{\delta}(q, g)$ | Upper bound of $\delta(q, g)$ |
| $f$ | (Partial) mapping from vertices of $q$ to vertices of $g$ |
| $f(v)$ | The vertex in $V(g)$ to which $v \in V(q)$ maps |
| $f^-(u)$ | The vertex in $V(q)$ that maps to $u \in V(g)$ |
| $\delta_f(q, g)$ | The editorial cost of the full mapping $f$ from $V(q)$ to $V(g)$ |
| $\underline{\delta}(f)$ | Lower bound of editorial costs of full mappings that extend $f$ |
| $\mathcal{T}$ | The search tree of all vertex mappings from $V(q)$ to $V(g)$ |
| $\mathcal{T}_{\leq \delta(q,g)}$ | The set of partial mappings $f$ in $\mathcal{T}$ s.t. $\underline{\delta}(f) \leq \delta(q, g)$ |
| $q[f]$ | The subgraph of $q$ induced by vertices of $q$ that are in $f$ |
| $q \backslash f$ | The remaining subgraph of $q$ by removing $q[f]$ |
| $\underline{\delta}(q \backslash f, g \backslash f)$ | Lower bound cost of mapping $q \backslash f$ to $g \backslash f$ |
| $\Upsilon(S_1, S_2)$ | Edit distance between multisets, i.e., $\max\{|S_1|, |S_2|\} - |S_1 \cap S_2|$ |

operations that transform $q$ to be isomorphic to $g$, given that $|V(q)| \leq |V(g)|$.

**Lemma 2.1:** *Given graphs $q$ and $g$ with $|V(q)| \leq |V(g)|$, there is no vertex deletion in the optimal sequence (i.e., with the minimum number) of edit operations that transform $q$ to be isomorphic to $g$.*

**Proof:** We prove the lemma by contradiction. Assume there is such a sequence of edit operations, $P = (eo_1, \ldots, eo_i, \ldots, eo_n)$ with length $n = \delta(q, g)$, that contains a vertex deletion. Without loss of generality, let $eo_i$ be the operation of deleting a vertex $v$ from $q$. Then, there must also exist a vertex insertion operation since $|V(q)| \leq |V(g)|$; let $eo_j$ be the operation of inserting a vertex $v'$ with label $a$, and $j$ can be either smaller or larger than $i$. Now consider another sequence $P'$ of edit operations that differs from $P$ by removing $eo_i$ and changing $eo_j$ from vertex insertion to vertex relabeling (*i.e.*, change the label of $v$ to $a$); note that, we also need to replace $v'$, occurred in $P$, with $v$. It is easy to verify that $P'$ also transforms $q$ to be isomorphic to $g$, and $|P'| = |P| - 1$, which contradicts that $P$ is optimal. Thus, the lemma holds. □

It is easy to verify that if $|V(q)| < |V(g)|$, then $\delta(q, g) = \delta(q', g)$ where $q'$ is obtained from $q$ by adding $|V(g)| - |V(q)|$ isolated vertices with the unique label $\perp \notin \Sigma$. Moreover, similar to Lemma 2.1, we can prove that if $|V(q)| = |V(g)|$, then there is no vertex deletion nor vertex insertion in the optimal sequence of edit operations that transform $q$ to be isomorphic to $g$. Thus, in the following, *for presentation simplicity we assume $q$ and $g$ have the same number of vertices, and we only need to consider four edit operations (i.e., edge insertion/deletion, and vertex/edge relabeling)*; nevertheless, we have $|V(q)| \neq |V(g)|$ in our experiments.

**Compute GED via Vertex Mapping.** As vertex insertion and vertex deletion are not allowed in the edit operations, there is a natural *one-to-one mapping* from $V(q)$ to $V(g)$ that is preserved in the final isomorphism between the transformed graph $q'$, obtained from $q$ by applying the edit operations, and $g$. For presentation simplicity, we refer to one-to-one mapping as mapping in the following.

It has been shown in [13] that the GED between two graphs can be computed via enumerating vertex mappings, as follows.

*Definition 2.3:* Given a mapping $f$ from $V(q)$ to $V(g)$, the **ed-**

**Algorithm 1:** EditorialCost

**Input**: Graphs $q$ and $g$, and a mapping $f$ from $V(q)$ to $V(g)$
**Output**: Editorial cost $\delta_f(q, g)$

1 $\delta_f(q, g) \leftarrow 0$;
  /* Vertex relabeling                               */
2 **for each** vertex $v$ in $q$ **do**
3     **if** $l(v) \neq l(f(v))$ **then** $\delta_f(q, g) \leftarrow \delta_f(q, g) + 1$;
  /* Edge deletion or relabeling                     */
4 **for each** edge $(v, v')$ in $q$ **do**
5     **if** edge $(f(v), f(v')) \notin g$ **or** $l(v, v') \neq l(f(v), f(v'))$ **then**
6        $\delta_f(q, g) \leftarrow \delta_f(q, g) + 1$;
  /* Edge insertion                                  */
7 **for each** edge $(u, u')$ in $g$ **do**
8     **if** edge $(f^-(u), f^-(u')) \notin q$ **then** $\delta_f(q, g) \leftarrow \delta_f(q, g) + 1$;
9 **return** $\delta_f(q, g)$;

itorial cost of $f$, denoted by $\delta_f(q, g)$, is the minimum number of edit operations that is required to transform $q$ to be isomorphic to $g$ by obeying the mapping $f$ (i.e., $v \in V(q)$ maps to $f(v) \in V(g)$ in the isomorphism).

**Lemma 2.2:** *[13] The GED between $q$ and $g$ equals the minimum editorial cost among all mappings from $V(q)$ to $V(g)$; that is, $\delta(q, g) = \min_{f \in \mathcal{F}(q,g)} \delta_f(q, g)$, where $\mathcal{F}(q, g)$ denotes all mappings from $V(q)$ to $V(g)$.*

Thus, the GED between $q$ and $g$ can be computed by enumerating all mappings from $V(q)$ to $V(g)$ and computing their editorial costs, and then the minimum editorial cost is the result. The editorial cost of a mapping from $V(q)$ to $V(g)$ can be computed in $O(\text{size}(q) + \text{size}(g))$ time, where such an algorithm is shown in Algorithm 1. Note that, this does not contradict with the NP-hardness of computing GED, since there is an exponential number of mappings from $V(q)$ to $V(g)$.

## 3 A Unified Framework

From Section 2.1, we know that the GED between two graphs $q$ and $g$ can be computed by enumerating vertex mappings from $V(q)$ to $V(g)$, rather than enumerating sequences of edit operations which has a much larger search space. However, there is still an exponential number (specifically, $|V(g)|!$) of mappings. In this section, we develop a unified framework to conduct an efficient *pruned search* on the search tree, which compactly represents the exponential number of mappings from $V(q)$ to $V(g)$, for finding the mapping with the minimum editorial cost.

**Search Tree.** Given graphs $q$ and $g$, and a matching order $\pi = (v_1, \ldots, v_{|V(q)|})$ of $V(q)$, we compactly represent all vertex mappings from $V(q)$ to $V(g)$, according to the matching order $\pi$, in a prefix-shared *search tree* $\mathcal{T}$. A node at level $i$ (from the root) of $\mathcal{T}$ represents a partial mapping from $(v_1, \ldots, v_i)$ to $V(g)$, which extends that of its parent at level $i - 1$ by also mapping $v_i$ to a vertex in $V(g)$. To distinguish the nodes in the search tree $\mathcal{T}$ from vertices in graphs $q$ and $g$, we refer to the former as nodes. We use the term "mapping" when the context applies to both partial mapping and full mapping; otherwise, we explicitly specify "partial mapping" or "full mapping". Based on the parent-child relationships of nodes in $\mathcal{T}$, parent-child (as well as ancestor-descendant) relationships are also defined for mappings in $\mathcal{T}$. We call a mapping $f$ *extending* a partial mapping $f'$ if $f$ is a descendant of $f'$; equivalently, each $v \mapsto u$ (i.e., $v$ maps to $u$) of $f'$ is also in $f$.

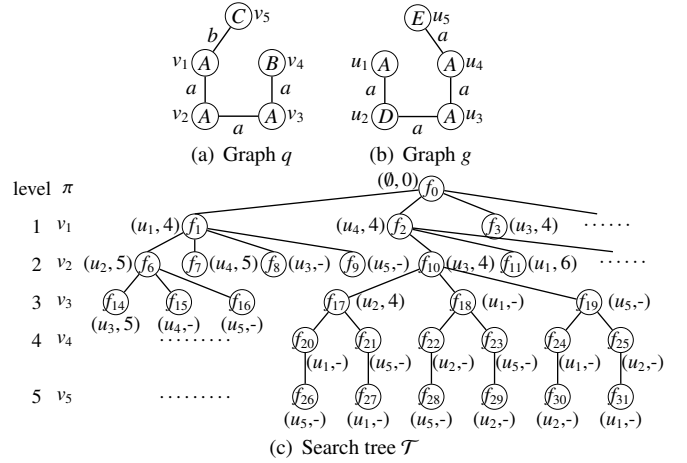

Figure 4: The search tree $\mathcal{T}$ for computing $\delta(q, g)$: $f_i$ is a partial mapping, and beside $f$ at level $j$ is a pair $(u, \delta(f))$ where $u \in V(g)$ is the vertex to which $v_j$ maps

A snippet of the search tree $\mathcal{T}$ of all vertex mappings from the graph $q$ in Figure 4(a) to the graph $g$ in Figure 4(b) is shown in Figure 4(c). The root node of $\mathcal{T}$ is at level 0, and represents an empty mapping. Beside each node $f$ at level $i$ in $\mathcal{T}$, we show two values: the vertex in $V(g)$ to which $v_i$ maps in the mapping $f$, and the lower bound cost of $f$ which will be introduced shortly; note that, we use $f$ to denote both a mapping and its corresponding node in $\mathcal{T}$. The vertex to which $v_j$ maps in the mapping $f$ for $j < i$ can be obtained from the corresponding ancestor of $f$ at level $j$ in $\mathcal{T}$. The full mappings from $V(q)$ to $V(g)$ are at level $|V(q)|$ of $\mathcal{T}$. For example, the partial mapping $f_1$ is $\{v_1 \mapsto u_1\}$ and has a lower bound cost 4, while the partial mapping $f_2$ is $\{v_1 \mapsto u_1, v_2 \mapsto u_2\}$ and has a lower bound cost 5.

**The Framework.** To avoid enumerating all mappings for computing $\delta(q, g)$, a lower bound cost is computed for each partial mapping for the purpose of pruning.

*Definition 3.1:* The **lower bound cost of a mapping** $f$, denoted by $\delta(f)$, is a value that is no larger than the minimum editorial cost among all full mappings that extend $f$.

As a result, we can safely prune all mappings that extend a partial mapping $f$ if $\delta(f)$ is already no smaller than the minimum editorial cost among all currently enumerated full mappings. To do so, we maintain the minimum editorial cost among all enumerated full mappings as the *upper bound* $\overline{\delta}(q, g)$ of $\delta(q, g)$. If the lower bound cost $\delta(f)$ of a partial mapping $f$ is no smaller than $\overline{\delta}(q, g)$, then we terminate the search on the subtree of $\mathcal{T}$ rooted at $f$ by not extending $f$. Based on this idea, the pseudocode of computing $\delta(q, g)$ by conducting a pruned search on $\mathcal{T}$ is shown in Algorithm 2.

We first compute a matching order of vertices of $q$ and let it be $\pi = (v_1, \ldots, v_{|V(q)|})$ (Line 1), initialize the upper bound

**Algorithm 2:** A Unified Framework

**Input**: Graphs $q$ and $g$
**Output**: GED between $q$ and $g$: $\delta(q,g)$

1 Compute a matching order $\pi = (v_1, \ldots, v_{|V(q)|})$ of vertices of $q$;
2 $\overline{\delta}(q,g) \leftarrow +\infty$;   /* Set the upper bound */;
3 Initialize a priority queue $Q$ with an empty mapping $(\emptyset, 0, 0, \emptyset)$;
4 **while** $Q \neq \emptyset$ **do**
5  | Pop $(f, i, \delta(f), C(v_i))$ from $Q$;
6  | **if** $\delta(f) < \overline{\delta}(q,g)$ **then**
   |   | /* Generate $f$'s next best sibling */
7  |   | **if** $i > 0$ **and** $C(v_i) \neq \emptyset$ **then** GenNext$(f, i, C(v_i))$;
   |   | /* Generate $f$'s best child */
8  |   | **if** $i < |V(q)|$ **then** GenNext$(f, i+1, V(g\backslash f))$;
9 **return** $\overline{\delta}(q,g)$;

**Procedure** GenNext(partial mapping $f$, level $i$, candidates $C(v_i)$)
10 $u^* \leftarrow \arg\min_{u \in C(v_i)} \delta(f \cup \{v_i \mapsto u\})$;
11 **if** $\delta(f \cup \{v_i \mapsto u^*\}) < \overline{\delta}(q,g)$ **then**
12  | Push $(f \cup \{v_i \mapsto u^*\}, i, \delta(f \cup \{v_i \mapsto u^*\}), C(v_i)\backslash u^*)$ into $Q$;
13  | Generate a full mapping $f'$ by extending $f \cup \{v_i \mapsto u^*\}$, and update $\overline{\delta}(q,g)$ based on $\delta_{f'}(q,g)$;   /* Optional */;

$\overline{\delta}(q,g)$ (Line 2), and initialize a priority queue $Q$ which stores all mappings to be extended (Line 3). Each entry of the priority queue stores a mapping $f$, its level $i$ in $\mathcal{T}$, its lower bound cost $\delta(f)$, and also the set $C(v_i)$ of candidates to which $v_i$ can map in the ungenerated siblings of $f$; here, the siblings of $f$ are the mappings in $\mathcal{T}$ that share the same parent as $f$. In this way, *the ungenerated siblings of $f$, each of which has a lower bound cost at least $\delta(f)$, are compactly represented by $C(v_i)$*. Then, we iteratively pop an entry $(f, i, \delta(f), C(v_i))$ from $Q$ (Line 5). If the lower bound cost $\delta(f)$ is at least $\overline{\delta}(q,g)$, then we do not extend $f$ (Line 6), which essentially prunes the subtrees of $\mathcal{T}$ rooted at $f$ and rooted at $f$'s ungenerated siblings (Line 6). Otherwise, we generate the best ungenerated sibling (Line 7) and the best child (Line 8) of $f$ in $\mathcal{T}$, regarding the lower bound cost, and push them into $Q$ (Lines 11–12); *in this case, we call $f$ as being extended*. Here, $V(g\backslash f)$ denotes the set of vertices of $g$ that are not used in $f$. At Line 7, in computing the best ungenerated sibling of $f$, we first remove the current mapping of $v_i$ from $f$; thus, the best sibling and the best child of $f$ are computed by the same procedure GenNext, which finds *the best extension of $f$ by additionally mapping $v_i$*. For a generated partial mapping $f$, we may optionally construct a full mapping $f'$ that extends $f$, and update the upper bound $\overline{\delta}(q,g)$ by $\delta_{f'}(q,g)$ (Line 13).

**Generality of the Framework.** The framework in Algorithm 2 is general and can be instantiated into different algorithms as follows. (1) At Line 5, priority queues with different strategies for the popping operation can be adopted, based on which Algorithm 2 can be instantiated into either a best-first search approach (see Section 5.1), or a depth-first search approach (see Section 5.2). (2) At Lines 10–11, different lower bound estimation techniques can be adopted (see Section 4). (3) At Line 1, different matching orders of vertices of $q$ can be obtained (see Section A.1 in Appendix). Moreover, all instantiations of our general framework in Algorithm 2 correctly compute the GED between two graphs, as proved below.

**Theorem 3.1:** *Algorithm 2 correctly computes the GED between two graphs.*

**Proof:** We prove the theorem by contradiction. Assume Algorithm 2 outputs a wrong value for $\delta(q,g)$. Then, when the algorithm terminates, we must have $\overline{\delta}(q,g) > \delta(q,g)$ and $Q = \emptyset$. From Lemma 2.2, we know that $\delta(q,g)$ equals the minimum editorial cost among all full mappings from $V(q)$ to $V(g)$; without loss of generality, we assume that there is only one full mapping $f^*$ with editorial cost $\delta(q,g)$. Then, at some point, an ancestor of $f^*$ in the search tree $\mathcal{T}$ must be pruned by the algorithm, since the empty mapping $f_0$ which is an ancestor of $f^*$ is initially contained in $Q$. A partial mapping $f$ is pruned (at Lines 6 and 11) only if $\delta(f) \geq \overline{\delta}(q,g)$. However, all ancestors of $f^*$ have lower bound costs $\leq \delta(q,g)$, and $\overline{\delta}(q,g) > \delta(q,g)$. We reach a contradiction. Thus, the theorem holds. □

## 4 Anchor-aware Lower Bounds

In this section, we develop anchor-aware lower bound estimation techniques to compute a tight lower bound for partial mappings. We first present the lower bounds in Section 4.1, and then propose techniques to efficiently compute the best extension of a partial mapping regarding the lower bounds in Section 4.2.

### 4.1 Lower Bounds

**Lower Bound Estimation Framework.** For a given partial mapping $f$, we denote the subgraph of $q$ (resp. $g$) induced by vertices in $f$ as $q[f]$ (resp. $g[f]$), and denote the remaining subgraph of $q$ (resp. $g$) as $q\backslash f$ (resp. $g\backslash f$). Note that, $q\backslash f$ contains none of the vertices of $q[f]$ but includes edges that have exactly one end-point in $q[f]$. Thus, $q\backslash f$ contains both *inner edges* whose both end-points are in $q\backslash f$, and *cross edges* between vertices of $q\backslash f$ and vertices of $q[f]$. This also holds for $g\backslash f$.

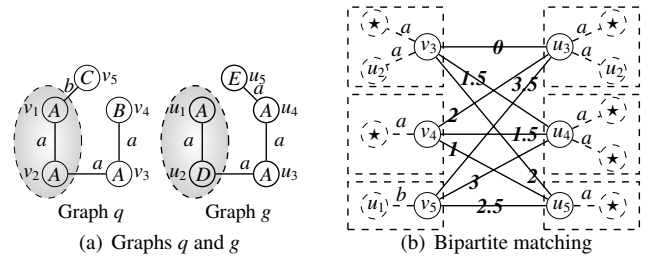

(a) Graphs $q$ and $g$    (b) Bipartite matching
Figure 5: Graphs and bipartite matching

**Example 4.1:** Consider the partial mapping $f = \{v_1 \mapsto u_1, v_2 \mapsto u_2\}$ for the graphs $q$ and $g$ in Figure 5(a). $q[f]$ and $g[f]$ are the parts in the shadowed area, while $q\backslash f$ and $g\backslash f$ are the remaining parts; specifically, $q\backslash f$ consists of three vertices $\{v_3, v_4, v_5\}$, one inner edge $\{(v_3, v_4)\}$ and two cross edges $\{(v_5, v_1), (v_3, v_2)\}$. □

Based on $q[f], g[f], q\backslash f$ and $g\backslash f$, we decompose the lower bound cost $\delta(f)$ of a partial mapping $f$ into two parts, (1) the cost to transform $q[f]$ to be isomorphic to $g[f]$ by obeying the mapping $f$, and (2) a lower bound cost for editing vertices and edges of $q\backslash f$ to be isomorphic to $g\backslash f$. It is easy to see that the first part is exactly $\delta_f(q[f], g[f])$ which can be computed in linear time (see Section 2.1). We denote the second part as $\underline{\delta}(q\backslash f, g\backslash f)$, whose computation is the focus of this section. Thus, $\delta(f) := \delta_f(q[f], g[f]) + \underline{\delta}(q\backslash f, g\backslash f)$.

**Existing Lower Bounds.** The existing techniques for computing a *global lower bound* of $\delta(q, g)$ are mainly label set-based [5, 23, 24], branch match-based [31], and star match-based [28]. Note that, *while the first one has been used for estimating $\underline{\delta}(q\backslash f, g\backslash f)$, the latter two have not been utilized in the context of GED computation or verification partially due to their cubic computational cost.* In the following, we illustrate the first two for estimating $\underline{\delta}(q\backslash f, g\backslash f)$. The star match-based lower bound is inferior to the branch match-based lower bound, and is given in Section A.3 in Appendix.

*Label Set-based Lower Bound $\underline{\delta}^{\text{LS}}(\cdot,\cdot)$.* Let $L_V(q\backslash f)$ and $L_V(g\backslash f)$ denote the multi-sets of vertex labels of $q\backslash f$ and $g\backslash f$, respectively, and $L_E(q\backslash f)$ and $L_E(g\backslash f)$ denote the multi-sets of edge labels. The label set-based lower bound [5] is,

$$\underline{\delta}^{\text{LS}}(q\backslash f, g\backslash f) := \Upsilon(L_V(q\backslash f), L_V(g\backslash f)) + \Upsilon(L_E(q\backslash f), L_E(g\backslash f)),$$

where $\Upsilon(\cdot, \cdot)$ denotes the edit distance between two multi-sets and $\Upsilon(S_1, S_2) = \max\{|S_1|, |S_2|\} - |S_1 \cap S_2|$ for multi-sets $S_1$ and $S_2$ (please see Section A.2 in Appendix for details). For the partial mapping $f$ in Example 4.1, $L_V(q\backslash f) = \{A, B, C\}$, $L_V(g\backslash f) = \{A, A, E\}$, $L_E(q\backslash f) = \{a, a, b\}$, and $L_E(g\backslash f) = \{a, a, a\}$. Thus, $\underline{\delta}^{\text{LS}}(q\backslash f, g\backslash f) = 2 + 1 = 3$.

*Branch Match-based Lower Bound $\underline{\delta}^{\text{BM}}(\cdot, \cdot)$.* Lower bound based on the branch structure is proposed in [31].

*Definition 4.1:* The **branch** of a vertex $v$ is $B(v) = (l(v), L_E(v))$, where $L_E(v)$ denotes the multi-set of labels of $v$'s adjacent edges.

Based on the branch structures $B(v)$ and $B(u)$, the *mapping cost* of mapping $v \in q\backslash f$ to $u \in g\backslash f$ is defined as,

$$\lambda^{\text{BM}}(v, u) := \mathbb{1}_{l(v) \neq l(u)} + \tfrac{1}{2} \times \Upsilon(L_E(v), L_E(u)),$$

where $\mathbb{1}_\phi$ is an indicator function that equals 1 if the expression $\phi$ evaluates true and 0 otherwise. Then, the branch match-based lower bound [31] is,

$$\underline{\delta}^{\text{BM}}(q\backslash f, g\backslash f) := \min_{\sigma \in \mathcal{F}(q\backslash f, g\backslash f)} \sum_{v \in q\backslash f} \lambda^{\text{BM}}(v, \sigma(v)),$$

where $\mathcal{F}(q\backslash f, g\backslash f)$ denotes the set of all mappings from vertices of $q\backslash f$ to vertices of $g\backslash f$. For the partial mapping $f$ in Example 4.1, $B(v_3) = (A, \{a, a\})$ and $B(u_4) = (A, \{a, a\})$; thus, $\lambda^{\text{BM}}(v_3, u_4) = 0$. It can be verified that $\underline{\delta}^{\text{BM}}(q\backslash f, g\backslash f) = 3$.

**Our Anchor-aware Lower Bounds.** The above existing lower bounds for estimating $\underline{\delta}(q\backslash f, g\backslash f)$ are loose, since they do not exploit the mapping information of the partial mapping $f$. In the following, we propose anchor-aware techniques to improve $\underline{\delta}^{\text{LS}}(\cdot, \cdot)$ and $\underline{\delta}^{\text{BM}}(\cdot, \cdot)$.

*Definition 4.2:* We call vertices of $q$ in $f$ as **anchored vertices** since their mappings are fixed, and vertices of $q$ not in $f$ as **free vertices** since they can freely map to any vertex in $g\backslash f$.

*Anchor-aware Label Set-based Lower Bound $\underline{\delta}^{\text{LSa}}(\cdot, \cdot)$.* As each anchored vertex $v$ in $q[f]$ is fixed to map to $f(v)$, the set of adjacent cross edges of $v$ must be edited to map to that of $f(v)$ in every full mapping that extends $f$. Let $L_{E_I}(q\backslash f)$ and $L_{E_I}(g\backslash f)$ be the multi-sets of labels of inner edges of $q\backslash f$ and $g\backslash f$, respectively, and $L_{E_C}(v)$ be the multi-set of labels of $v$'s adjacent cross edges. It can be verified that $L_{E_I}(q\backslash f)$ and $L_{E_C}(v)$ for anchored vertices $v$ in $q[f]$ refine $L_E(q\backslash f)$; that is, $L_E(q\backslash f) = L_{E_I}(q\backslash f) \cup (\bigcup_{v \in q[f]} L_{E_C}(v))$. Then, we define the anchor-aware label set-based lower bound as,

$$\underline{\delta}^{\text{LSa}}(q\backslash f, g\backslash f) := \Upsilon(L_V(q\backslash f), L_V(g\backslash f)) + \Upsilon(L_{E_I}(q\backslash f), L_{E_I}(g\backslash f)) + \sum_{v \in q[f]} \Upsilon(L_{E_C}(v), L_{E_C}(f(v))),$$

whose correctness can be easily verified.

It can also be verified from the above and Lemma A.1 in Appendix that $\underline{\delta}^{\text{LSa}}(q\backslash f, g\backslash f) \geq \underline{\delta}^{\text{LS}}(q\backslash f, g\backslash f)$. Note that, for lower bounds, the larger the better. For the partial mapping $f$ in Example 4.1, we have $L_{E_C}(v_1) = \{b\}$, $L_{E_C}(v_2) = \{a\}$, $L_{E_C}(u_1) = \emptyset$ and $L_{E_C}(u_2) = \{a\}$. Thus, $\underline{\delta}^{\text{LSa}}(q\backslash f, g\backslash f) = 4$.

*Anchor-aware Branch Match-based Lower Bound $\underline{\delta}^{\text{BMa}}(\cdot, \cdot)$.* To exploit the anchored vertices for the branch match-based lower bound $\underline{\delta}^{\text{BM}}(\cdot, \cdot)$, we revise the branch structure of a vertex $v \in q\backslash f$ as $B'(v) = (l(v), L_{E_I}(v), \bigcup_{v' \in q[f]} \{(f(v'), l(v, v'))\})$, where $l(v, v') = \bot$ if the edge is not in $q$. That is, we explicitly consider each anchored vertex $v'$ and its connection $l(v, v')$ to $v$. Similarly, we revise the branch structure of a vertex $u \in g\backslash f$ as $B'(u) = (l(u), L_{E_I}(u), \bigcup_{u' \in g[f]} \{(u', l(u, u'))\})$. The mapping cost of mapping $v \in q\backslash f$ to $u \in g\backslash f$ is revised to be the sum of the edit distances between the three corresponding components of $B'(v)$ and $B'(u)$, that is,

$$\lambda^{\text{BMa}}(v, u) := \mathbb{1}_{l(v) \neq l(u)} + \tfrac{1}{2} \times \Upsilon(L_{E_I}(v), L_{E_I}(u)) + \sum_{v' \in q[f]} \mathbb{1}_{l(v, v') \neq l(u, f(v'))}.$$

where the label of a non-existent edge is defined as $\bot$. Intuitively, $\lambda^{\text{BMa}}(v, u)$ equals the minimum cost to edit $v$ and its adjacent edges (*i.e.*, the branch structure of $v$) to be the same as $u$ and $u$'s adjacent edges, subject to the constraint that an adjacent edge of $v$ connecting to an anchored vertex $v'$ must map to the adjacent edge of $u$ connecting to $f(v')$. For example, in Figure 5(b), the structure in each dotted rectangle is a branch, where non-existence edges (*i.e.*, with label $\bot$) are omitted; ★ denotes a free vertex and can map to any free vertex. The mapping cost between two vertices are illustrated, in the middle part, on the solid edge connecting the vertices. In particular, we have $B'(v_3) = (A, \{a\}, \{(u_1, \bot), (u_2, a)\})$ and $B'(u_4) = (A, \{a, a\}, \{(u_1, \bot), (u_2, \bot)\})$; thus, $\lambda^{\text{BMa}}(v_3, u_4) = 1.5$.

Then, we define the anchor-aware branch match-based lower bound as,

$$\underline{\delta}^{\text{BMa}}(q\backslash f, g\backslash f) := \min_{\sigma \in \mathcal{F}(q\backslash f, g\backslash f)} \sum_{v \in q\backslash f} \lambda^{\text{BMa}}(v, \sigma(v)),$$

whose correctness can be proved in a similar way to the proof of $\underline{\delta}^{\text{BM}}(q\backslash f, g\backslash f)$ in [31]. It can be easily verified that $\lambda^{\text{BMa}}(v, u) \geq \lambda^{\text{BM}}(v, u)$. Consequently, $\underline{\delta}^{\text{BMa}}(q\backslash f, g\backslash f) \geq \underline{\delta}^{\text{BM}}(q\backslash f, g\backslash f)$. For the partial mapping $f$ in Example 4.1, based on the mapping costs as shown in Figure 5(b), we have $\underline{\delta}^{\text{BMa}}(q\backslash f, g\backslash f) = 4$.

We prove in the lemma below that $\underline{\delta}^{\text{BMa}}(\cdot, \cdot)$ computes a tighter lower bound than $\underline{\delta}^{\text{LSa}}(\cdot, \cdot)$.

**Lemma 4.1:** *For any partial mapping $f$, we have $\underline{\delta}^{\text{BMa}}(q\backslash f, g\backslash f) \geq \underline{\delta}^{\text{LSa}}(q\backslash f, g\backslash f)$.*

**Proof:** It suffices to prove that for any mapping $\sigma \in \mathcal{F}(q\backslash f, g\backslash f)$, the following holds: $\sum_{v \in q\backslash f} \lambda^{\text{BMa}}(v, \sigma(v)) \geq \underline{\delta}^{\text{LSa}}(q\backslash f, g\backslash f) = \Upsilon(L_V(q\backslash f), L_V(g\backslash f)) + \Upsilon(L_{E_I}(q\backslash f), L_{E_I}(g\backslash f)) + \sum_{v \in q[f]} \Upsilon(L_{E_C}(v), L_{E_C}(f(v)))$.

Firstly, we have,

$$\sum_{v \in q\backslash f} \lambda^{\text{BMa}}(v, \sigma(v))$$
$$= \sum_{v \in q\backslash f} \left( \mathbb{1}_{l(v) \neq l(\sigma(v))} + \tfrac{1}{2} \times \Upsilon(L_{E_I}(v), L_{E_I}(\sigma(v))) \right.$$
$$\left. + \sum_{v' \in q[f]} \mathbb{1}_{l(v,v') \neq l(\sigma(v), f(v'))} \right)$$
$$= \sum_{v \in q\backslash f} \mathbb{1}_{l(v) \neq l(\sigma(v))} + \tfrac{1}{2} \times \sum_{v \in q\backslash f} \Upsilon(L_{E_I}(v), L_{E_I}(\sigma(v)))$$
$$+ \sum_{v \in q\backslash f} \sum_{v' \in q[f]} \mathbb{1}_{l(v,v') \neq l(\sigma(v), f(v'))}.$$

Secondly, it is easy to verify the following three inequalities.

**(1)** $\sum_{v \in q\backslash f} \mathbb{1}_{l(v) \neq l(\sigma(v))} \geq \Upsilon(L_V(q\backslash f), L_V(g\backslash f))$

since $L_V(q\backslash f) = \bigcup_{v \in q\backslash f} \{l(v)\}$ and $L_V(g\backslash f) = \bigcup_{v \in q\backslash f} \{l(\sigma(v))\}$.

**(2)** $\tfrac{1}{2} \times \sum_{v \in q\backslash f} \Upsilon(L_{E_I}(v), L_{E_I}(\sigma(v))) \geq \Upsilon(L_{E_I}(q\backslash f), L_{E_I}(g\backslash f))$

since $L_{E_I}(q\backslash f) \cup L_{E_I}(q\backslash f) = \bigcup_{v \in q\backslash f} L_{E_I}(v)$ and similar for $L_{E_I}(g\backslash f)$.

**(3)** $\sum_{v \in q\backslash f} \sum_{v' \in q[f]} \mathbb{1}_{l(v,v') \neq l(\sigma(v), f(v'))}$
$\geq \sum_{v \in q[f]} \Upsilon(L_{E_C}(v), L_{E_C}(f(v)))$

since $\sum_{v' \in q\backslash f} \mathbb{1}_{l(v', v) \neq l(\sigma(v'), f(v))} \geq \Upsilon(L_{E_C}(v), L_{E_C}(f(v)))$ and

$$\sum_{v \in q\backslash f} \sum_{v' \in q[f]} \mathbb{1}_{l(v,v') \neq l(\sigma(v), f(v'))}$$
$$= \sum_{v \in q[f]} \sum_{v' \in q\backslash f} \mathbb{1}_{l(v',v) \neq l(\sigma(v'), f(v))}.$$

Thus, the lemma holds. □

*Remark.* The lower bound estimation technique in CSI_GED [11] also exploits the anchored vertices; we denote it by $\underline{\delta}^{\text{DEa}}(\cdot, \cdot)$ since it only uses the **d**egree information. Specifically, (1) regarding cross edges that are adjacent to anchored vertices, $\underline{\delta}^{\text{DEa}}(\cdot, \cdot)$ considers only the difference between the numbers of adjacent cross edges of mapped anchored vertices while ignoring edge labels. (2) Regarding inner edges, $\underline{\delta}^{\text{DEa}}(\cdot, \cdot)$ considers only the difference between the numbers of edges while ignoring edge labels. Consequently, we have $\underline{\delta}^{\text{LSa}}(q\backslash f, g\backslash f) \geq \underline{\delta}^{\text{DEa}}(q\backslash f, g\backslash f)$. Moreover, our experimental results show that our depth-first search approach incorporated with our $\underline{\delta}^{\text{LSa}}(\cdot, \cdot)$ lower bound significantly outperforms CSI_GED (see Section 6). Thus, we omit $\underline{\delta}^{\text{DEa}}(\cdot, \cdot)$ in this paper.

## 4.2 Efficiently Compute the Best Extension

In Section 4.1, we have illustrated four lower bound estimation techniques for computing $\underline{\delta}(q\backslash f, g\backslash f)$. For each of them, we can define a lower bound cost of $f$ as $\delta(f) := \delta_f(q[f], g[f]) + \underline{\delta}(q\backslash f, g\backslash f)$. Thus, we have lower bounds $\delta^{\text{LS}}(\cdot), \delta^{\text{BM}}(\cdot), \delta^{\text{LSa}}(\cdot), \delta^{\text{BMa}}(\cdot)$. In Algorithm 2, given a partial mapping $f$ and a lower bound definition $\delta(\cdot)$, we need to compute the best extension of $f$; that is, compute $\arg\min_{u \in C(v_i)} \delta(f \cup \{v_i \mapsto u\})$, where $v_i \notin f$ is the next vertex of $q$ to map according to the mapping order $\pi$ and $C(v_i)$ is the subset vertices of $g$ that $v_i$ can map to. In this subsection, we propose techniques to efficiently compute the best extension of $f$ regarding the lower bounds.

**Compute the Best Extension of $f$ regarding $\delta^{\text{BM}}(\cdot)$ and $\delta^{\text{BMa}}(\cdot)$.** These two lower bounds are similar to each other by only differing in the mapping cost $\lambda(v, u)$, which can be computed in $O(d(v) + d(u))$ time. Thus, in the following we mainly focus our discussions on $\delta^{\text{BMa}}(\cdot)$, while the other is omitted.

Firstly, given a partial mapping $f$, we can compute $\underline{\delta}^{\text{BMa}}(q\backslash f, g\backslash f) = \min_{\sigma \in \mathcal{F}(q\backslash f, g\backslash f)} \sum_{v \in q\backslash f} \lambda^{\text{BMa}}(v, \sigma(v))$ in $O((|V(q)| + |V(g)|)^3)$ time, by the classic Hungarian algorithm [16, 19] which computes the minimum cost perfect matching in a bipartite graph [9]. Here, the bipartite graph consists of vertices of $q\backslash f$ on one side and vertices of $g\backslash f$ on the other side, where each edge $(v, u)$ has a cost $\lambda^{\text{BMa}}(v, u)$. For example, for the partial mapping $f$ in Example 4.1, the bipartite graph is constructed as illustrated in Figure 5(b), where edges of the bipartite graph are shown as solid lines with their costs shown on the edges. By running the Hungarian algorithm, we obtain the minimum cost perfect matching as $\{v_3 \mapsto u_3, v_4 \mapsto u_4, v_5 \mapsto u_5\}$ whose cost is 4; thus, $\underline{\delta}^{\text{BMa}}(q\backslash f, g\backslash f) = 4$.

For a partial mapping $f'$, we can define a naive lower bound as $\delta^{\text{BMaN}}(f') := \delta_{f'}(q[f'], g[f']) + \underline{\delta}^{\text{BMa}}(q\backslash f', g\backslash f')$. However, computing the best extension of a partial mapping $f$ regarding $\delta^{\text{BMaN}}(\cdot)$ needs $O(|V(g)| \times (|V(q)| + |V(g)|)^3)$ time. This is because, we need to iterate through every candidate $u \in C(v_i)$ of $v_i$ and compute its lower bound cost $\delta^{\text{BMaN}}(f \cup \{v_i \mapsto u\})$. Note that, the costs of edges in the bipartite graph constructed for computing $\delta^{\text{BMaN}}(f \cup \{v_i \mapsto u\})$ differ a lot from that constructed for computing $\delta^{\text{BMaN}}(f \cup \{v_i \mapsto u'\})$; they can be completely different in the worst case. As a result, computations may not be shared when computing the lower bounds of different extensions of $f$.

---

**Algorithm 3:** Compute best extension of $f$ regarding $\delta^{\text{BMa}}(\cdot)$

**Input**: Graphs $q$ and $g$, a partial mapping $f$, and candidates $C(v_i)$ of the next mapping vertex $v_i \in q\backslash f$
**Output**: Best extension $u^* \leftarrow \arg\min_{u \in C(v_i)} \delta^{\text{BMa}}(f \cup \{v_i \mapsto u\})$

1 **for each** vertex $v$ in $q\backslash f$ **do**
2     **for each** vertex $u$ in $g\backslash f$ **do**
3         **if** $v = v_i$ **and** $u \notin C(v_i)$ **then** $\lambda(v, u) \leftarrow +\infty$;
4         **else** Compute the mapping cost $\lambda(v, u)$ of mapping $v$ to $u$;

5 Compute the minimum cost perfect matching $M$ between vertices of $q\backslash f$ and vertices of $g\backslash f$ by invoking the Hungarian method in [16, 19];
6 Let $u^*$ be the vertex to which $v_i$ maps in $M$;
7 **return** $u^*$;

In view of the above, we propose to define the lower bound as $\delta^{\text{BMa}}(f \cup \{v_i \mapsto u\}) := \delta_f(q[f], g[f]) + \underline{\delta}^{\text{BMa}}_{v_i \mapsto u}(q\backslash f, g\backslash f)$ where $\underline{\delta}^{\text{BMa}}_{v_i \mapsto u}(q\backslash f, g\backslash f)$ is similar to $\underline{\delta}^{\text{BMa}}(q\backslash f, g\backslash f)$ but $v_i$ is constrained to map to $u$. Moreover, in computing $\underline{\delta}^{\text{BMa}}_{v_i \mapsto u}(q\backslash f, g\backslash f)$, $v_i$ is considered as a free rather than anchored vertex. Recall that, $v_i$ is considered as an anchored vertex in computing $\underline{\delta}^{\text{BMaN}}(q\backslash (f \cup \{v_i \mapsto u\}), g\backslash (f \cup \{v_i \mapsto u\}))$. Thus, intuitively $\delta^{\text{BMa}}(f \cup \{v_i \mapsto u\}) \leq \delta^{\text{BMaN}}(f \cup \{v_i \mapsto u\})$; nevertheless, the gap between these two lower bounds are not large as will be illustrated in Section 5.1. The advantage of $\delta^{\text{BMa}}(\cdot)$ is that the best extension of a partial mapping regarding $\delta^{\text{BMa}}(\cdot)$ can be computed in $O((|V(q)| + |V(g)|)^3)$ time, by conducting only one computation of the minimum cost perfect matching, based on the property that $\min_{u \in C(v_i)} \underline{\delta}^{\text{BMa}}_{v_i \mapsto u}(q\backslash f, g\backslash f) = \underline{\delta}^{\text{BMa}}_{v_i \mapsto C(v_i)}(q\backslash f, g\backslash f)$. The pseudocode is shown in Algorithm 3, which is self-explanatory. We prove its correctness in below.

**Theorem 4.1:** *Algorithm 3 correctly computes the best extension of a partial mapping $f$ regarding $\delta^{\text{BMa}}(\cdot)$, where the next vertex $v_i$ can map to a vertex from $C(v_i)$.*

**Proof:** It is easy to see that, if $C(v_i)$ consists of only one vertex (i.e., $C(v_i) = \{u^*\}$), then Algorithm 3 correctly maps $v_i$ to $u^*$ and moreover the cost of the matching $M$ equals $\underline{\delta}^{\text{BMa}}_{v_i \mapsto u^*}(q\backslash f, g\backslash f)$. Now, we prove the theorem by contradiction. Assume the best extension of $f$ is $\{v_i \mapsto u'\}$; that is, $\underline{\delta}^{\text{BMa}}_{v_i \mapsto u'}(q\backslash f, g\backslash f) < \underline{\delta}^{\text{BMa}}_{v_i \mapsto u^*}(q\backslash f, g\backslash f)$. Then, the computation of $\underline{\delta}^{\text{BMa}}_{v_i \mapsto u'}(q\backslash f, g\backslash f)$ also implies a bipartite matching, which exists in the bipartite graph constructed by Algorithm 3 and has a smaller cost than $M$. This contradicts that $M$ is the minimum cost perfect matching. Thus, the theorem holds. □

Note that, we can also extend Algorithm 3 to compute the lower bound costs of every extension of $f$ in $O((|V(q)| + |V(g)|)^3)$ total time. That is, after computing a minimum cost perfect matching $M$, we change the cost $\lambda(v_i, u')$ for the mapping $v_i \mapsto u'$ in $M$ into $+\infty$, and then extend $M\backslash \{v_i \mapsto u'\}$ into a minimum cost perfect matching in $O((|V(q)| + |V(g)|)^2)$ time.

**Compute the Best Extension of $f$ regarding $\delta^{\text{LS}}(\cdot)$ and $\delta^{\text{LSa}}(\cdot)$.** Given a partial mapping $f'$, we define $\delta^{\text{LS}}(f') := \delta_{f'}(q[f'], g[f']) + \underline{\delta}^{\text{LS}}(q\backslash f', g\backslash f')$. This actually is the lower bound definition used in the existing best-first search algorithm A*GED [23, 24, 30]. However, A*GED uses a naive approach to computing the best extension of $f$ regarding $\delta^{\text{LS}}(\cdot)$; that is, compute $\delta^{\text{LS}}(f \cup \{v_i \mapsto u\})$ independently for each vertex $u \in C(v_i)$, and then choose the one that results in the minimum lower bound. As each lower bound is computed in $O(\text{size}(q) + \text{size}(g))$ time, this naive approach has a time complexity of $O(|V(g)| \times (\text{size}(q) + \text{size}(g)))$.

For efficient GED computation and verification, we propose an algorithm to compute the best extension of $f$ regarding $\delta^{\text{LS}}(\cdot)$ in $O(\text{size}(q) + \text{size}(g))$ time, which also computes $\delta^{\text{LS}}(f \cup \{v_i \mapsto u\})$ for each $u$ in $C(v_i)$. The pseudocode is shown in Algorithm 4. The general idea of our algorithm is that after constructing some data structures in $O(\text{size}(q)+\text{size}(g))$ time (Lines 1–7), we compute $\delta^{\text{LS}}(f \cup \{v_i \mapsto u\})$ in $O(d(u))$ time for each $u \in C(v_i)$ (Lines 10–23), where $d(u)$ is the degree of $u$ in $g$. Specifically, we let $f'$ denote $f \cup \{v_i \mapsto u\}$, and we illustrate how

---

**Algorithm 4:** Compute best extension of $f$ regarding $\delta^{\text{LS}}(\cdot)$

**Input**: Graphs $q$ and $g$, a partial mapping $f$, and candidates $C(v_i)$ of the next mapping vertex $v_i \in q\backslash f$
**Output**: Best extension $u^* \leftarrow \arg\min_{u \in C(v_i)} \delta^{\text{LS}}(f \cup \{v_i \mapsto u\})$

1  Let $d_1$ be the number of edges between $v_i$ and vertices of $q[f]$;
2  Let $n_1$ and $n_2$ be the numbers of edges in $q\backslash (f \cup \{v_i\})$ and $g\backslash f$, respectively;
3  Let $n_E(a)$ be the number of edges in $g\backslash f$ with label $a$ minus the number of edges in $q\backslash (f \cup \{v_i\})$ with label $a$ for each edge label $a$ appeared in $g\backslash f$;
4  Let $c_E$ be the cardinality of the multi-set intersection between the edge labels of $q\backslash (f \cup \{v_i\})$ and the edge labels of $g\backslash f$;
5  Let $n_V(A)$ be the number of vertices in $g\backslash f$ with label $A$ minus the number of vertices in $q\backslash (f \cup \{v_i\})$ with label $A$ for each vertex label $A$ appeared in $g\backslash f$;
6  Let $c_V$ be the cardinality of the multi-set intersection between the vertex labels of $q\backslash (f \cup \{v_i\})$ and the vertex labels of $g\backslash f$;
7  $max_V \leftarrow \max\{|V(g\backslash f)| - 1, |V(g\backslash f)| - 1\}$;
8  $u^* \leftarrow \text{null}$;
9  **for each** vertex $u$ in $C(v_i)$ **do**
10    $d_2 \leftarrow 0, c_1 \leftarrow 0, c_2 \leftarrow 0$;
11    **for each** edge $(u, u')$ between $u$ and $g[f]$ **do**
12      $d_2 \leftarrow d_2 + 1$;
13      **if** edge $(v_i, f^-(u'))$ exists in $q$ **then**
14        **if** $l(u, u') = l(v_i, f^-(u'))$ **then** $c_2 \leftarrow c_2 + 1$;
15        **else** $c_1 \leftarrow c_1 + 1$;
16      $n_2 \leftarrow n_2 - 1, a \leftarrow l(u, u'), n_E(a) \leftarrow n_E(a) - 1$;
17      **if** $n_E(a) < 0$ **then** $c_E \leftarrow c_E - 1$;
18    $\delta^{\text{LS}}(f \cup \{v_i \mapsto u\}) \leftarrow \delta_f(q[f], g[f]) + (d_1 + d_2 - 2 \times c_2 - c_1 + \mathbb{1}_{l(v_i) \neq l(u)}) + ((\max\{n_1, n_2\} - c_E) + (max_V - c_V + \mathbb{1}_{n_V(l(u)) \leq 0}))$;
19    **if** $u^* = \text{null}$ **or** $\delta^{\text{LS}}(f \cup \{v_i \mapsto u\}) < \delta^{\text{LS}}(f \cup \{v_i \mapsto u^*\})$ **then**
20      $u^* \leftarrow u$;
      /* Restore $n_2, n_E(a), c_E$           */
21    **for each** edge $(u, u')$ between $u$ and $g[f]$ **do**
22      $n_2 \leftarrow n_2 + 1, a \leftarrow l(u, u'), n_E(a) \leftarrow n_E(a) + 1$;
23      **if** $n_E(a) \leq 0$ **then** $c_E \leftarrow c_E + 1$;
24  **return** $u^*$;

---

to compute $\delta^{\text{LS}}(f') = \delta_{f'}(q[f'], g[f']) + \underline{\delta}^{\text{LS}}(q\backslash f', g\backslash f')$. Let $(v_i, q[f])$ and $(u, g[f])$ denote the edges between $v_i$ and $q[f]$, and the edges between $u$ and $g[f]$, respectively. Firstly, we have $\delta_{f'}(q[f'], g[f']) = \delta_f(q[f], g[f]) + (d_1 + d_2 - 2 \times c_2 - c_1 + \mathbb{1}_{l(v_i) \neq l(u)})$, where $d_1$ and $d_2$ are the numbers of edges in $(v_i, q[f])$ and $(u, g[f])$, respectively, $c_2$ is the number of matched edges with the same labels between $(v_i, q[f])$ and $(u, g[f])$ (i.e., no edit operation is required), and $c_1$ is the number of matched edges with different labels between $(v_i, q[f])$ and $(u, g[f])$ (i.e., edge relabeling is required). It can be verified that the second part of this equation equals the minimum cost to edit the edges in $(v_i, q[f])$ to map to the edges in $(u, g[f])$ according to the mapping $f'$.

Secondly, we have $\underline{\delta}^{\text{LS}}(q\backslash f', g\backslash f') = \Upsilon(L_V(q\backslash f'), L_V(g\backslash f')) + \Upsilon(L_E(q\backslash f'), L_E(g\backslash f'))$. In the following, we discuss how to compute $\Upsilon(L_E(q\backslash f'), L_E(g\backslash f'))$, while $\Upsilon(L_V(q\backslash f'), L_V(g\backslash f'))$ can be computed similarly. We compute it as $\max\{n_1, n_2\} - c_E$, where $n_1 = |L_E(q\backslash f')|$, $n_2 = |L_E(g\backslash f')|$ and $c_E = |L_E(q\backslash f') \cap L_E(g\backslash f')|$. $n_1$ is computed in the initialization step (Line 2). $n_2$ is computed as $|L_E(g\backslash f)| - |(u, g[f])|$, where $|L_E(g\backslash f)|$ is computed in the initialization step and $|(u, g[f])|$ is computed on-demand in $O(d(u))$ time. $c_E$ is computed based on $|L_E(q\backslash f') \cap L_E(g\backslash f)|$ that is obtained in the initialization step, and the surplus data structure $n_E(a)$ which equals the number of edges in $g\backslash f$ with label $a$ minus the number of edges in $q\backslash f'$ with label $a$. Note that, given two multi-sets $S_1$ and $S_2$, and one element $a \in S_2$, we have $|S_1 \cap S_2| = |S_1 \cap S_2| - 1$ if and only if the surplus $n_E(a)$ before removing $a$ from $S_2$ is non-positive (*i.e.*, $S_2$ has no more $a$'s than $S_1$). Thus, Algorithm 4 correctly computes the best extension of $f$ and also the lower bound cost of every extension of $f$, regarding $\delta^{LS}(\cdot)$, in $O(\texttt{size}(q) + \texttt{size}(g))$ total time.

Similarly, we define $\delta^{LSa}(f') := \delta_{f'}(q[f'], g[f']) + \underline{\delta}^{LSa}(q\backslash f', g\backslash f')$. The best extension of $f$ as well as the lower bound cost of every extension of $f$, regarding $\delta^{LSa}(\cdot)$, can be computed in $O(\texttt{size}(q) + \texttt{size}(g))$ total time in a similar, but more involved, way to Algorithm 4. We omit the details.

**Remark on Maintaining an Upper Bound.** In Algorithm 2, we maintain an upper bound $\bar{\delta}(q, g)$, which equals the minimum editorial cost among all enumerated full mappings, to reduce the memory consumption in practice. Thus, after generating a new partial mapping at Line 12 of Algorithm 2, we can heuristically extend it to a full mapping for updating $\bar{\delta}(q, g)$. It is easy to see that the minimum cost perfect matching $M$ computed by Algorithm 3 actually extends $f$ to a full mapping. Thus, we use the editorial cost of the full mapping $f \cup M$ to update $\bar{\delta}(q, g)$, when lower bound $\delta^{BM}(\cdot)$ or $\delta^{BMa}(\cdot)$ is used. On the other hand, we do not heuristically extend $f$ to a full mapping if lower bound $\delta^{LS}(\cdot)$ or $\delta^{LSa}(\cdot)$ is used.

## 5 Our Best-First and Depth-First Search Approaches

In this section, we first instantiate our framework (Algorithm 2) into a best-first search approach AStar$^+$ in Section 5.1, and a depth-first search approach DFS$^+$ in Section 5.2. Then, we contrast these two approaches for the problems of GED computation and GED verification in Section 5.3.

### 5.1 Our AStar$^+$ Approach

Our framework in Algorithm 2 works in a best-first search fashion if at Line 5, we pop from the priority queue $Q$ the partial mapping with the minimum lower bound cost; if there is a tie on the minimum lower bound cost, then we break the tie by preferring large level numbers. Note that, if a tie still occurs, then we break the tie arbitrarily. We denote our approach with this strategy as AStar$^+$.

**Running Example of AStar$^+$.** Consider the graphs $q$ and $g$ in Figures 4(a) and 4(b), respectively. Assume the matching order in computing $\delta(q, g)$ is $\pi = (v_1, v_2, v_3, v_4, v_5)$, and the lower bound costs of partial mappings are as shown in Figure 4(c) which are actually computed by $\delta^{BMa}(\cdot)$. AStar$^+$ runs as follows. Initially, $Q = \{f_0\}, \bar{\delta}(q, g) = +\infty$. In the first iteration, we pop $f_0$ from $Q$, and push its best child $f_1$ into $Q$. Assume the full mapping obtained by extending $f_1$ is $(u_1, u_4, u_3, u_5, u_2)$ which has an editorial cost 7, we update $\bar{\delta}(q, g)$ as 7. Thus, $Q = \{f_1\}$ and $\bar{\delta}(q, g) = 7$. In the second iteration, we pop $f_1$ from $Q$, and push both its best ungenerated sibling $f_2$ and its best child $f_6$ into $Q$. The full mappings that extend $f_2$ and $f_6$ for updating the upper bound are $(u_4, u_1, u_3, u_5, u_2)$ and $(u_1, u_2, u_3, u_5, u_4)$, respectively; the upper bound $\bar{\delta}(q, g)$ remains 7. Thus, $Q = \{f_2, f_6\}$ and $\bar{\delta}(q, g) = 7$. In the third iteration, we pop $f_2$ from $Q$ and push both its best ungenerated sibling $f_3$ and its best child $f_{10}$ into $Q$. The full mappings that extend $f_3$ and $f_{10}$ for updating the upper bound are $(u_3, u_1, u_4, u_5, u_2)$ and $(u_4, u_3, u_1, u_2, u_5)$, respectively; the upper bound $\bar{\delta}(q, g)$ is updated as 5. Thus, $Q = \{f_6, f_3, f_{10}\}$ and $\bar{\delta}(q, g) = 5$. In the fourth iteration, we pop $f_{10}$ from $Q$ and push its best child $f_{17}$ into $Q$; note that, its best ungenerated sibling $f_{11}$ is not pushed into $Q$ since its lower bound cost is larger than $\bar{\delta}(q, g)$. The full mapping that extends $f_{17}$ for updating the upper bound is $(u_4, u_3, u_2, u_1, u_5)$, which updates $\bar{\delta}(q, g)$ as 4. Thus, $Q = \{f_6, f_3, f_{17}\}$ and $\bar{\delta}(q, g) = 4$. Then, partial mappings are iteratively popped from $Q$ without being extended since their lower bound costs are no smaller than $\bar{\delta}(q, g)$. As a result, the GED between $q$ and $g$ is reported as 4.

**Analysis of AStar$^+$.** Let $\mathcal{T}_{\leq \delta(q,g)}$ be the set of non-leaf nodes (*i.e.*, partial mappings) in $\mathcal{T}$ whose lower bound costs are no larger than $\delta(q, g)$, and $|\mathcal{T}_{\leq \delta(q,g)}|$ be its cardinality. Here, we assume that *the sequence of lower bound costs for any root to leaf path in $\mathcal{T}$ is non-decreasing*. Note that, in Algorithm 2, it is possible that when extending a partial mapping $f$ to get its best ungenerated sibling $f'$ (Line 7) and its best child $f''$ (Line 8), the computed lower bound cost of $f'$ (or $f''$) is smaller than that of $f$; nevertheless, if this is the case, then we can safely change the lower bound cost of $f'$ (or $f''$) to be that of $f$. We have the following lemma.

**Lemma 5.1:** *The set of partial mappings extended by AStar$^+$ (i.e., reach Lines 7–8 of Algorithm 2) is a subset of $\mathcal{T}_{\leq \delta(q,g)}$.*

**Proof:** Note that, a partial mapping is extended only after being popped from the priority queue $Q$. Consider the moment of popping from the priority queue $Q$ a partial mapping $f$ whose lower bound cost is larger than $\delta(q, g)$. The lower bound costs of all remaining partial mappings in $Q$ are no smaller than the lower bound cost $\delta(f)$ of $f$, based on our strategy of popping from $Q$ the partial mapping with the minimum lower bound cost. Then, the upper bound $\bar{\delta}(q, g)$ must equal $\delta(q, g)$; otherwise, an ancestor $f'$ of the minimum-cost full mapping $f^*$ (*i.e.*, with editorial cost $\delta(q, g)$) must be in $Q$ since the ancestor $f_0$ (*i.e.*, the empty mapping) is initially in $Q$, which would contradict that $f$ is popped from $Q$ since $\delta(f') \leq \delta(q, g) < \delta(f)$. Thus, $f$ will not be extended, and the lemma holds. □

Intuitively, the tighter the lower bound, the smaller the $|\mathcal{T}_{\leq \delta(q,g)}|$. In the following, we use the superscript to denote the lower bound used. Thus, we have $|\mathcal{T}^{BMa}_{\leq \delta(q,g)}| \leq |\mathcal{T}^{LSa}_{\leq \delta(q,g)}| \leq |\mathcal{T}^{LS}_{\leq \delta(q,g)}|$, and $|\mathcal{T}^{BMa}_{\leq \delta(q,g)}| \leq |\mathcal{T}^{BM}_{\leq \delta(q,g)}|$. Note that, although $|\mathcal{T}^{BMa}_{\leq \delta(q,g)}| \geq |\mathcal{T}^{BMaN}_{\leq \delta(q,g)}|$, they do not differ too much as proved by the lemma below.

**Lemma 5.2:** $|\mathcal{T}^{\text{BMa}}_{\leq \delta(q,g)}| \leq |V(g)| \times |\mathcal{T}^{\text{BMaN}}_{\leq \delta(q,g)}|$.

**Proof:** It is easy to see that $\delta^{\text{BMa}}(f_2) \geq \delta^{\text{BMaN}}(f_1)$ holds for each child $f_2$ of $f_1$ in the search tree $\mathcal{T}$; note that, for lower bounds, the larger the better. As a result, for each partial mapping $f \notin \mathcal{T}^{\text{BMaN}}_{\leq \delta(q,g)}$, all its children are not in $\mathcal{T}^{\text{BMa}}_{\leq \delta(q,g)}$. Thus, the lemma holds. □

In practice, $|\mathcal{T}^{\text{BMa}}_{\leq \delta(q,g)}|$ is much smaller than $|V(g)| \times |\mathcal{T}^{\text{BMaN}}_{\leq \delta(q,g)}|$, and is similar to $|\mathcal{T}^{\text{BMaN}}_{\leq \delta(q,g)}|$.

*Space Complexity of* AStar$^+$. The space complexity of AStar$^+$ is $O(|\mathcal{T}_{\leq \delta(q,g)}|)$. As the space consumption is mainly dominated by the priority queue $Q$, we first bound the number of partial mappings that were pushed into $Q$, as follows.

**Lemma 5.3:** *The total number of partial mappings that were pushed into the priority queue $Q$ when running* AStar$^+$ *is bounded by* $O(|\mathcal{T}_{\leq \delta(q,g)}|)$.

**Proof:** Firstly, we prove that the number of partial mappings in $Q$ after $i$ iterations is $\leq i$ by induction, where an iteration is running once Lines 4–8 of Algorithm 2. Initially, the claim holds for the first iteration, since the empty mapping has no sibling; thus, we pop the empty mapping from $Q$ and push its best child into $Q$. Assume the claim holds for $i \geq 1$, we prove that it also holds for $i+1$. At the $(i+1)$-th iteration, we pop one partial mapping from $Q$ and push at most two partial mappings into $Q$ (one at Line 7 and another at Line 8); the number of partial mappings in $Q$ increases by at most 1 and thus becomes $\leq i+1$. Therefore, the claim holds for all iterations.

Secondly, following from Lemma 5.1, the number of iterations before popping a partial mapping $f$ with $\delta(f) \geq \overline{\delta}(q,g)$ is at most $|\mathcal{T}_{\leq \delta(q,g)}|$. Moreover, after popping the first partial mapping with lower bound cost $\geq \overline{\delta}(q,g)$, no new partial mappings will be generated or pushed into $Q$. Thus, the lemma holds. □

Although $Q$ may also include partial mappings whose lower bound costs are larger than $\delta(q,g)$, the number of partial mappings that were ever pushed into $Q$ is still bounded by $O(|\mathcal{T}_{\leq \delta(q,g)}|)$. Now, we prove the space complexity of AStar$^+$ by the following theorem.

**Theorem 5.1:** *The space complexity of* AStar$^+$ *is* $O(|\mathcal{T}_{\leq \delta(q,g)}|)$.

**Proof:** *The information of each partial mapping (i.e., each entry) in $Q$ can be stored in constant space* as follows. (1) For a partial mapping $f$ at level $i$, we only store the vertex of $V(g)$ to which $v_i \in V(q)$ maps, while other parts of $f$ can be retrieved from its ancestors in the search tree $\mathcal{T}$. (2) The level number and the lower bound cost of $f$ also take constant spaces. (3) We store a pointer for $f$ pointing to its immediate preceding sibling, from which $f$ is generated; thus, $C(v_i)$ as needed at Line 7 can be retrieved online in $O(|C(v_i)|)$ time. As a result, after a partial mapping being popped from $Q$, we do not remove it from the main memory. Nevertheless, the space complexity of AStar$^+$ is still bounded by $O(|\mathcal{T}_{\leq \delta(q,g)}|)$. □

In practice, we further reduce the space consumption of AStar$^+$ based on the maintained upper bound $\overline{\delta}(q,g)$. Specifically, 1) we remove from $Q$ all partial mappings whose lower bound costs are no smaller than $\overline{\delta}(q,g)$, and 2) we also remove from the main memory the partial mappings that have no descendent in $Q$.

*Time Complexity of* AStar$^+$. Let $T_{\text{BE}}$ be the time of computing the best extension of a partial mapping at Line 10 of Algorithm 2, which depends on the actual lower bound estimation technique and is discussed in Section 4.2. The time complexity of AStar$^+$ is $O(|\mathcal{T}_{\leq \delta(q,g)}| \times (\log |\mathcal{T}_{\leq \delta(q,g)}| + T_{\text{BE}}))$, as proved in the following theorem.

**Theorem 5.2:** *The time complexity of* AStar$^+$ *is* $O(|\mathcal{T}_{\leq \delta(q,g)}| \times (\log |\mathcal{T}_{\leq \delta(q,g)}| + T_{\text{BE}}))$.

**Proof:** Firstly, following Lemma 5.3, AStar$^+$ runs for at most $O(|\mathcal{T}_{\leq \delta(q,g)}|)$ iterations; that is, *the search space of* AStar$^+$ *is* $O(|\mathcal{T}_{\leq \delta(q,g)}|)$. Secondly, each iteration consists of one pop operation (Line 5) and at most two push operations (Line 12) of $Q$, each with time complexity $O(\log |\mathcal{T}_{\leq \delta(q,g)}|)$, and at most two computations of the best extension (Line 10), each with time complexity $T_{\text{BE}}$. Thus, the theorem holds. □

As $\log |\mathcal{T}_{\leq \delta(q,g)}|$ is bounded by $|V(q)| \times \log |V(g)|$ and is usually small than $T_{\text{BE}}$, we simply regard the time complexity of AStar$^+$ as $O(|\mathcal{T}_{\leq \delta(q,g)}| \times T_{\text{BE}})$.

**Expand All Strategy.** In the above discussions, to save memory space, each time when computing the best child of a partial mapping, we only materialize the best child and its lower bound cost. Then, to obtain the best ungenerated sibling of a partial mapping, we need to run the expensive best extension computation algorithm again (Line 7 of Algorithm 2). As a result, the same lower bound cost $\delta(f \cup \{v_i \mapsto u\})$ may be computed up-to $O(|V(g)|)$ times, once in computing the best sibling for each $f'$ that is a sibling of $f \cup \{v_i \mapsto u\}$; thus, there are redundant computations.

We can trade a little memory for time efficiency by the expand all strategy. That is, when computing the best child of a partial mapping, we generate all its children and compute their lower bound costs; note that, this still can be conducted in $T_{\text{BE}}$ time (see Section 4.2). Nevertheless, rather than pushing all these generated children into the priority queue $Q$, we only push the best child into $Q$ and attach all the remaining children to this best child. Subsequently, to obtain the best ungenerated sibling of a partial mapping, we directly select the best one that is attached to it.

**Comparison to the Existing Best-First Search Approach.** Although the best-first search strategy has been adopted in existing works (*e.g.,* A$^*$GED [23, 24, 29, 30]), *one unique feature of* AStar$^+$ *is reducing memory consumption* by storing each partial mapping in a constant memory space and utilizing the upper bound $\overline{\delta}(q,g)$ to prune the priority queue. Consequently, AStar$^+$ largely resolves the issue of running out-of-memory of the existing best-first search approach A$^*$GED. Moreover, we propose anchor-aware lower bounds and also efficient algorithms to compute the best extension of a partial mapping. In summary, A$^*$GED has both a larger space complexity of $O(|\mathcal{T}_{\leq \delta(q,g)}| \times |V(g)|)$ and a larger time complexity of $O(|\mathcal{T}_{\leq \delta(q,g)}| \times |V(g)| \times T_{\text{LB}})$, where $T_{\text{LB}}$ is the time to compute the lower bound cost of a partial mapping.

## 5.2 Our DFS$^+$ Approach

Algorithm 2 works in a depth-first search fashion if at Line 5, we pop from the priority queue $Q$ the partial mapping that has the largest level number according to the search tree $\mathcal{T}$. Then, at any time, there is at most one partial mapping kept in $Q$ for each distinct level number. Thus, we can simulate the priority queue by an array of size $|V(q)|$. We denote our algorithm with this strategy as DFS$^+$.

**Running Example of DFS$^+$.** Reconsider the above running example. Now, we run DFS$^+$ to compute the GED between $q$ and $g$ in Figures 4(a) and 4(b), respectively. The initialization and the first two iterations are the same as that of running AStar$^+$, and we have $Q = \{f_2, f_6\}$ and $\bar{\delta}(q, g) = 7$. In the third iteration, we pop $f_6$ from $Q$ since its level number is the largest, and push both its best ungenerated sibling $f_7$ and its best child $f_{14}$ into $Q$. The full mappings that extend $f_7$ and $f_{14}$ for updating the upper bound are $(u_1, u_4, u_3, u_5, u_2)$ and $(u_1, u_2, u_3, u_4, u_5)$, respectively; the upper bound $\bar{\delta}(q, g)$ is updated as 5. Thus $Q = \{f_2, f_7, f_{14}\}$ and $\bar{\delta}(q, g) = 5$. Then, $f_{14}$ and $f_7$ are popped from $Q$ without being extended in the fourth and fifth iterations, respectively, since their lower bound costs are no less than $\bar{\delta}(q, g)$. Thus $Q = \{f_2\}$ and $\bar{\delta}(q, g) = 5$. In the sixth iteration, We pop $f_2$ from $Q$ and update $Q$ and $\bar{\delta}(q, g)$ in a similar fashion to the third iteration of the running example of AStar$^+$. Thus $Q = \{f_3, f_{10}\}$ and $\bar{\delta}(q, g) = 5$. In the seventh iteration, we pop $f_{10}$ from $Q$ and update $Q$ and $\bar{\delta}(q, g)$ in a similar way to the fourth iteration of running AStar$^+$. Thus $Q = \{f_3, f_{17}\}$ and $\bar{\delta}(q, g) = 4$. Then, partial mappings $f_{17}$ and $f_3$ are iteratively popped from $Q$ without being extended since their lower bound costs are no smaller than $\bar{\delta}(q, g)$. As a result, the GED between $q$ and $g$ is reported as 4.

**Space Complexity of DFS$^+$.** The space complexity of DFS$^+$ is $O(|V(q)| \times |V(g)|)$, as follows. Firstly, the number of partial mappings in the priority queue $Q$ when running DFS$^+$ is $O(|V(q)|)$. Secondly, for each partial mapping, we need to store either $C(v_i)$ or its remaining siblings depending on whether the expand all strategy is used.

**Time Complexity of DFS$^+$.** Let $\mathcal{T}_{\text{DFS}}$ be the set of all nodes (i.e., partial mappings) in $\mathcal{T}$ that are extended by DFS$^+$ (i.e., reach Lines 7–8 of Algorithm 2). That is, *the search space of DFS$^+$ is $O(|\mathcal{T}_{\text{DFS}}|)$*. It is easy to verify that the time complexity of DFS$^+$ is $O(|\mathcal{T}_{\text{DFS}}| \times \text{T}_{\text{BE}})$, since it takes $O(\text{T}_{\text{BE}})$ time to find the best sibling/child of a partial mapping.

**Comparison to the Existing Depth-First Search Approaches.** The idea of using depth-first search for GED computation has also been exploited in the existing works; for example, DF_GED [1, 5] and CSI_GED [11]. DF_GED is similar to DFS$^+$ incorporated with the lower bound $\delta^{\text{LS}}(\cdot)$, but we propose a more efficient algorithm to compute the lower bound costs of all children of a partial mapping. On the other hand, CSI_GED enumerates edge mappings, while DFS$^+$ enumerates vertex mappings. Moreover, the lower bounds discussed in this paper, except $\delta^{\text{LS}}(\cdot)$, have not been used in the existing approaches.

## 5.3 AStar$^+$ v.s. DFS$^+$

Both AStar$^+$ and DFS$^+$ have merits and deficiencies. In the following, we analyze these two approaches for the problems of GED computation and GED verification, and suggest that AStar$^+$ is better than DFS$^+$ based on our tighter lower bound estimation techniques.

**GED Computation.** For the problem of GED computation, AStar$^+$ usually has a smaller search space (and thus lower time complexity), by extending fewer partial mappings, than DFS$^+$ based on the following intuitions. (1) Each non-leaf node of $\mathcal{T}$ that has a lower bound cost smaller than $\delta(q, g)$ must be extended by DFS$^+$; that is, $\mathcal{T}_{<\delta(q,g)} \subseteq \mathcal{T}_{\text{DFS}}$. This is because even if DFS$^+$ generates the minimum-cost full mapping $f^*$ at a very early stage of the searching, it still needs to exhaust all mappings in $\mathcal{T}_{<\delta(q,g)}$ to certify that there is no full mapping with a smaller editorial cost than $f^*$. (2) More often, DFS$^+$ also extends many partial mappings with lower bound costs larger than $\delta(q, g)$ due to a large $\bar{\delta}(q, g)$ obtained at early stages of the searching. Recall that, the time complexities of AStar$^+$ and DFS$^+$ are $O(|\mathcal{T}_{\leq\delta(q,g)}| \times \text{T}_{\text{BE}})$ and $O(|\mathcal{T}_{\text{DFS}}| \times \text{T}_{\text{BE}})$, respectively.

On the other hand, AStar$^+$ has a larger space complexity than DFS$^+$. One may think that it will run out-of-memory, as observed for the existing best-first search approach A*GED [1, 11]. However, AStar$^+$ largely resolves this issue by (1) representing each partial mapping by a constant memory space, (see Section 5.1), (2) incorporating upper bound to prune the priority queue (see Section 5.1), and (3) significantly reducing the search space $\mathcal{T}_{\leq\delta(q,g)}$ based on our anchor-aware tighter lower bounds (see Section 4). Consequently, AStar$^+$ *runs faster than DFS$^+$ as demonstrated by our experiments in Section 6, and thus is more suitable for GED computation.*

**GED Verification.** Our framework in Algorithm 2 can also process GED verification queries for a user-given threshold $\tau$, by a minor modification as follows. The upper bound $\bar{\delta}(q, g)$ is initialized to be $\tau + \epsilon$ at Line 2 for a small real value $\epsilon$ (e.g., 0.001), and the framework returns true once the full mapping generated at Line 13 has an editorial cost at most $\tau$. As a result, both AStar$^+$ and DFS$^+$ run faster for GED verification than for GED computation.

In the following, we show that AStar$^+$ *also suits for GED verification*. Firstly, if $q$ and $g$ are dissimilar (i.e., $\delta(q, g) > \tau$), then the sets of partial mappings in $\mathcal{T}$ that are extended by AStar$^+$ and DFS$^+$ are the same (i.e., $\mathcal{T}_{\leq\tau}$). Thus, AStar$^+$ and DFS$^+$ perform similarly for dissimilar queries. Secondly, if $q$ and $g$ are similar (i.e., $\delta(q, g) \leq \tau$), then both approaches can terminate early after finding a full mapping with editorial cost no larger than $\tau$, which although may not be the one with the minimum editorial cost. Nevertheless, AStar$^+$ has a higher chance of finding a full mapping with editorial cost no larger than $\tau$, due to the strategy of always extending the partial mapping with the minimum lower bound.

## 6 Experiments

We conduct extensive empirical studies to evaluate the effectiveness and efficiency of our techniques. To do so, we implemented several variants of our AStar$^+$ and DFS$^+$ algorithms by incorporating different lower bounds: AStar$^+$-LS, AStar$^+$-LSa, AStar$^+$-BM, AStar$^+$-BMa, AStar$^+$-BMaN, AStar$^+$-SMa, DFS$^+$-LSa and DFS$^+$-BMa;[2] for all these algorithms, we adopt the expand all strategy. All our algorithms are implemented in C++ and compiled with GNU GCC with the -O3 flag. We compare our algorithms with the existing best-first search approach A$^*$GED [24, 29], and the existing depth-first search approaches CSI_GED [11] and DF_GED [5]. Binary executable codes of A$^*$GED and CSI_GED are obtained from the authors of [29] and [11], respectively. We implemented DF_GED by our algorithm DFS$^+$-LS which actually improves DF_GED by computing lower bounds of partial mappings more efficiently. Note that, all existing works on graph similarity search use A$^*$GED for GED verification. All experiments are conducted on a machine with an Intel Xeon(R) 3.40GHz CPU and 16GB main memory.

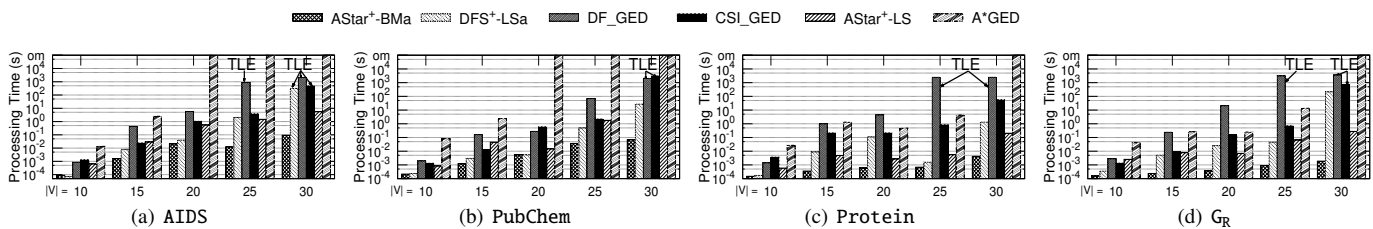

Figure 6: Against existing algorithms regarding processing time (vary |V|)

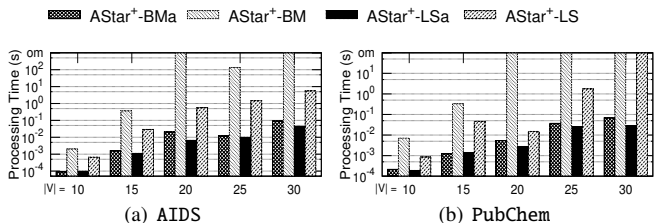

Figure 7: Evaluate anchor aware (processing time, vary |V|)

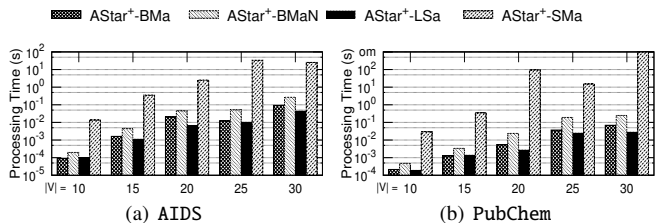

Figure 8: Evaluate lower bounds (processing time, vary |V|)

Table 3: Statistics of real graphs

| Graphs | Avg |V| | Avg |E| | #Vertex labels | #Edge labels |
|---|---|---|---|---|
| AIDS | 25.6 | 27.6 | 62 | 3 |
| PubChem | 24 | 25.8 | 81 | 3 |
| Protein | 32.6 | 62 | 3 | 5 |

**Real Graphs.** We evaluate the algorithms on three real graphs, AIDS, PubChem, and Protein, which are widely used in the existing works for graph similarity search [11, 17, 29, 30]. AIDS is an antivirus screen chemical compound dataset published by the Developmental Therapeutics Program at NCI/NIH[3]. It contains 42,687 chemical compounds with an average of 25.6 vertices and 27.6 edges. The numbers of distinct vertex labels and distinct edge labels are 62 and 3, respectively. PubChem is also a chemical compound dataset[4], and contains one million graphs with an average of 24 vertices and 25.8 edges. The numbers of distinct vertex labels and distinct edge labels are 81 and 3, respectively. Protein is a protein database from the Protein Data Bank[5], and consists of 600 protein structures with an average of 32.6 vertices and 62 edges. The number of distinct vertex labels and distinct edge labels are 3 and 5, respectively. Statistics of these real graphs are illustrated in Table 3.

**Synthetic Graphs.** We also generate synthetic random graphs $G_R$ by the graph generator GraphGen[6], to evaluate the algorithms. Specifically, we generate 10 groups of random graphs $G_R$, with the number of vertices chosen from $\{10, 15, 20, 25, 30, 64, 128, 256, 512, 1024\}$, where large graphs (i.e., with $|V| \geq 64$) are used for scalability testing and small graphs are used to compare the different algorithms. Each group of $G_R$ contains 51 graphs with the same number of vertices, and is generated as follows. We first generate a graph with $i$ vertices by invoking GraphGen, and then randomly apply $x$ edit operations on the graph 10 times to get 10 graphs, where $x$ is chosen from $\{1, 2, 3, 4, 5\}$ for small graphs and is chosen from $\{2, 5, 10, 20, 40\}$ for large graphs. Each graph generated by GraphGen has an edge density of 20%, 5 distinct vertex labels, and 2 distinct edge labels, similar to that used in [11].

**Generate Query Graph Pairs.** For each graph dataset and a specific number $i$ of vertices, we first select the graphs whose sizes are within the range of $[i - 2, i + 2]$, and then partition the set of all graph pairs among the selected graphs into different groups with respect to their GED values. Finally, 10 graph pairs are randomly sampled from each group, for which we run the algorithms. Thus, each reported experimental result for a specific graph size, and a specific GED value is an average of processing 10 graph pairs. We choose the group of graph pairs corresponding to GED = 9 by default.

**Evaluation Metrics.** For each testing, we report both the processing time and the search space. The search space of an algorithm is defined as the number of best extension computations, which is related to the space complexity $|\mathcal{T}_{\leq \delta(q,g)}|$ of AStar$^+$ algorithms. Each reported result is an average of processing 10 graph pairs in a group. We set a timeout of 1 hour (i.e., $3.6 \times 10^3$ seconds). If an algorithm takes more than 1 hour to process one graph pair, then we record this time as 1 hour and label the algorithm with "TLE" in the figure. We report the processing time and search space of an algorithm as "om" if it runs out-of-memory.

---

[2]A single binary executable code of all these algorithms can be downloaded from https://lijunchang.github.io/code/GED.zip
[3]http://dtp.nci.nih.gov/docs/aids/aids_data.html
[4]http://pubchem.ncbi.nlm.nih.gov
[5]http://www.iam.unibe.ch/fki/databases/iam-graph-database/download-the-iam-graph-database
[6]http://www.cse.cuhk.edu.hk/~jcheng/graphgen1.0.zip

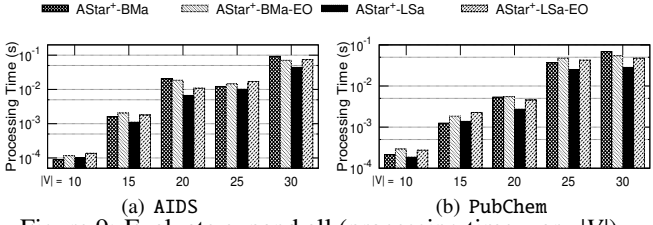

Figure 9: Evaluate expand all (processing time, vary |V|)

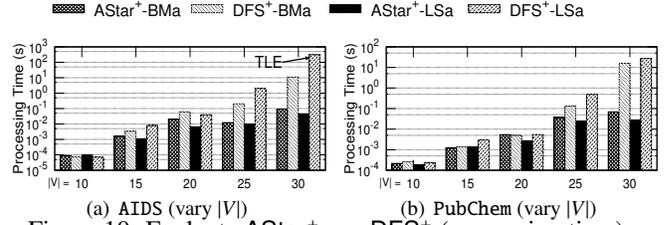

Figure 10: Evaluate AStar$^+$ v.s. DFS$^+$ (processing time)

## 6.1 Experimental Results for GED Computation

**Eval-I: Against Existing Algorithms.** We first evaluate our algorithms AStar$^+$-BMa, DFS$^+$-LSa and AStar$^+$-LS against the existing algorithms DF_GED, CSI_GED and A$^*$GED. The processing time of these algorithms on the four graphs by varying |V| is illustrated in Figure 6, where the query graph pairs have GED 9. We set the largest |V| as 30 because the existing algorithms fail to process graphs with 30 or more vertices. CSI_GED consistently performs better than A$^*$GED, which conforms with the observations in [11]. However, our AStar$^+$-LS algorithm, despite of adopting the same search strategy and the same lower bound as A$^*$GED, outperforms CSI_GED. This is because AStar$^+$-LS significantly improves upon A$^*$GED by (1) our strategy of reducing memory consumption which solves the out-of-memory issue of A$^*$GED (Section 5.1), and (2) our linear-time best extension computation algorithm which improves the time efficiency (Section 4.2). Moreover, as we illustrated in Section 5.3, best-first search has a much smaller search space and is more suitable than depth-first search for GED computation.

It is also interesting to observe that our depth-first algorithm DFS$^+$-LSa outperforms the two state-of-the-art depth-first algorithms CSI_GED and DF_GED. This is because our anchor-aware lower bound $\delta^{\text{LSa}}(\cdot)$ used in DFS$^+$-LSa is tighter than both the lower bound $\delta^{\text{LS}}(\cdot)$ used in DF_GED and the degree-based lower bound $\delta^{\text{DEa}}(\cdot)$ used in CSI_GED. That is, the tightness of a lower bound has a great impact on the performance of a GED computation algorithm. By incorporating a tighter lower bound $\delta^{\text{BMa}}(\cdot)$ and the better search strategy of AStar$^+$, AStar$^+$-BMa significantly outperforms all other algorithms, and the improvement of AStar$^+$-BMa over the state-of-the-art algorithms CSI_GED and DF_GED can be more than 4 orders of magnitude. For example, the average processing time of AStar$^+$-BMa on the PubChem graph with 30 vertices is less than 0.1 seconds, while both CSI_GED and DF_GED take more than $3 \times 10^3$ seconds.

**Eval-II: Evaluate the Effectiveness of Anchor Aware.** In this testing, we evaluate the effectiveness of our anchor-aware techniques for improving lower bound estimations. The results of comparing AStar$^+$-BMa, AStar$^+$-BM, AStar$^+$-LSa, and AStar$^+$-LS are shown in Figure 7. The AStar$^+$ algorithms with anchor-aware lower bounds (i.e., AStar$^+$-BMa and AStar$^+$-LSa) significantly outperform their baseline counterparts (i.e., AStar$^+$-BM and AStar$^+$-LS). This is because the anchor-aware techniques dramatically reduce the search space of an algorithm as shown in Figure 15 in Appendix, by computing tighter lower bounds. Thus, in the following, we only consider anchor-aware lower bounds.

**Eval-III: Evaluate Different Lower Bounds.** In this set of experiments, we evaluate the effect of different anchor-aware lower bounds within our AStar$^+$ algorithm. Specifically, we evaluate $\delta^{\text{BMa}}(\cdot), \delta^{\text{BMaN}}(\cdot), \delta^{\text{LSa}}(\cdot)$, and $\delta^{\text{SMa}}(\cdot)$. The results of processing time are shown in Figure 8, while that of search space are shown in Figure 16 in Appendix. The search spaces of the algorithms increase in the order of AStar$^+$-BMaN, AStar$^+$-BMa, AStar$^+$-LSa, AStar$^+$-SMa, because in general $\delta^{\text{BMaN}}(f) \geq \delta^{\text{BMa}}(f) \geq \delta^{\text{LSa}}(f) \geq \delta^{\text{SMa}}(f)$. When considering processing time, AStar$^+$-LSa runs the fastest due to our linear-time best extension computation algorithm regarding $\delta^{\text{LSa}}(\cdot)$. Nevertheless, the processing time difference between AStar$^+$-BMa and AStar$^+$-LSa is not significant. Recall that, the best extension of a partial mapping regarding $\delta^{\text{BMa}}(\cdot)$ and $\delta^{\text{SMa}}(\cdot)$ can be computed in cubic time, while that regarding $\delta^{\text{BMaN}}(\cdot)$ takes time to the power of four. Thus, there is a trade-off between the tightness of a lower bound estimation and its computational efficiency, and AStar$^+$-BMaN, despite of having the smallest search space, runs slower than AStar$^+$-BMa. In the following, we only consider the lower bounds $\delta^{\text{BMa}}(\cdot)$ and $\delta^{\text{LSa}}(\cdot)$.

**Eval-IV: Evaluate Expand All Strategy.** In this testing, we evaluate the effectiveness of the expand all strategy. We compare AStar$^+$-BMa and AStar$^+$-LSa with their counterparts AStar$^+$-BMa-EO and AStar$^+$-LSa-EO that only compute and materialize the best child when extending a partial mapping. The results are shown in Figure 9. We can see that the expand all strategy consistently improves AStar$^+$-LSa while having little effect on AStar$^+$-BMa. Thus, in the following, we adopt the expand all strategy.

**Eval-V: Evaluate AStar$^+$ Against DFS$^+$.** The results of evaluating the paradigms of AStar$^+$ and DFS$^+$ for GED computation are shown in Figure 10. We can see that AStar$^+$ consistently performs better than DFS$^+$, due to having a much smaller search space as shown in Figure 17 in Appendix. The improvement of AStar$^+$ over DFS$^+$ is more evident when the graph size increases. This confirms our recommendation that AStar$^+$ is better than DFS$^+$ for GED computation.

**Eval-VI: Scalability Testing of AStar$^+$-BMa and AStar$^+$-LSa.** We evaluate the scalability of AStar$^+$-BMa and AStar$^+$-LSa on AIDS and $G_R$ for different GED values by varying |V|. The largest AIDS graph has around 60 vertices, and we generate $G_R$ graphs with up-to 1024 vertices. The results in Figure 11 show that

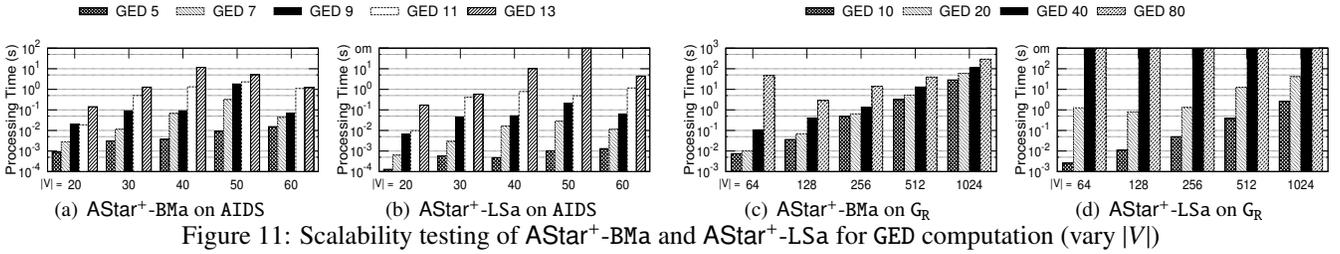

Figure 11: Scalability testing of $\mathsf{AStar}^+$-BMa and $\mathsf{AStar}^+$-LSa for GED computation (vary $|V|$)

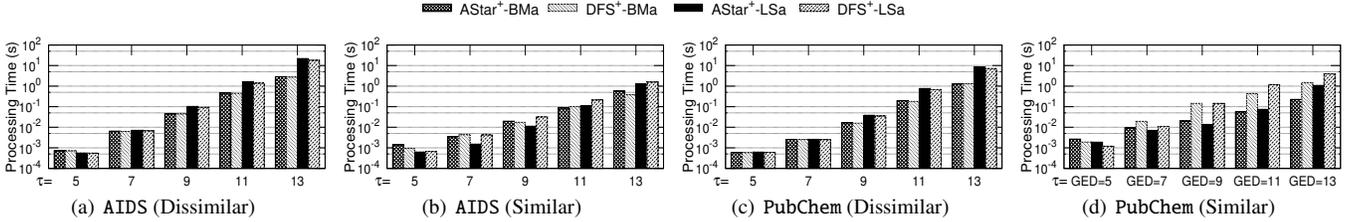

Figure 12: Evaluate $\mathsf{AStar}^+$ v.s. $\mathsf{DFS}^+$ for GED verification (vary $\tau$)

$\mathsf{AStar}^+$-BMa scales well for processing large graphs and large GED values, while $\mathsf{AStar}^+$-LSa may run out-of-memory due to having a larger search space (thus larger space complexity) as shown in Figure 18 in Appendix. Thus, $\mathsf{AStar}^+$-BMa is more suitable than $\mathsf{AStar}^+$-LSa for GED computation. Note that, $\mathsf{AStar}^+$-BMa manages to finish all the testings within the main memory limitation of our machine (*i.e.*, 16GB).

### 6.2 Experimental Results for GED Verification

**Eval-VII: Evaluate $\mathsf{AStar}^+$ Against $\mathsf{DFS}^+$.** In this testing, we evaluate the paradigms of $\mathsf{AStar}^+$ and $\mathsf{DFS}^+$ for GED verification, where the set of query graph pairs is obtained as the union of 10 pairs from each group with 30 vertices and corresponding to GED value in $\{5, 7, 9, 11, 13, \geq 14\}$. The results of processing time are shown in Figure 12, while that of search space are shown in Figure 19 in Appendix. We can see that, for dissimilar graph pairs, $\mathsf{AStar}^+$ and $\mathsf{DFS}^+$ perform similarly by having the same search space. For similar graph pairs, $\mathsf{AStar}^+$ runs slightly faster than $\mathsf{DFS}^+$ by having a smaller search space; nevertheless, the difference is not as significant as for GED computation in Figure 10. From Figure 12 we can also see that the lower bound $\delta^{\mathsf{BMa}}(\cdot)$ performs better than $\delta^{\mathsf{LSa}}(\cdot)$ for GED verification. Thus, in the following, we only consider $\delta^{\mathsf{BMa}}(\cdot)$.

**Eval-VIII: Against Existing Algorithms.** The results of evaluating $\mathsf{AStar}^+$-BMa and $\mathsf{DFS}^+$-BMa over CSI_GED and $\mathsf{AStar}^+$-LS are shown in Figure 13. Both $\mathsf{AStar}^+$-BMa and $\mathsf{DFS}^+$-BMa significantly outperform CSI_GED and $\mathsf{AStar}^+$-LS. Recall that, (1) $\mathsf{AStar}^+$-LS is an improved version of $\mathsf{A}^*\mathsf{GED}$ and performs better than $\mathsf{A}^*\mathsf{GED}$, and (2) $\mathsf{A}^*\mathsf{GED}$ is adopted for GED verification in the existing works on graph similarity search.

**Eval-IX: Scalability Testing of $\mathsf{AStar}^+$-BMa and $\mathsf{DFS}^+$-BMa.** The results of evaluating the scalability of $\mathsf{AStar}^+$-BMa and $\mathsf{DFS}^+$-BMa for GED verification are shown in Figure 14. We can see that, both algorithms scale well to process large graphs, and $\mathsf{AStar}^+$-BMa scale better for GED verification than for GED computation in Figure 11.

## 7 conclusion

In this paper, we developed a unified framework for GED computation and verification, which can be instantiated into either $\mathsf{AStar}^+$ or $\mathsf{DFS}^+$ search strategies. To speed up the computation, we further proposed anchor-aware tighter lower bound estimation techniques, as well as efficient techniques for computing the best extension of a partial mapping regarding a lower bound. Extensive performance studies confirm our theoretical analysis of the superiority of $\mathsf{AStar}^+$ over $\mathsf{DFS}^+$. Moreover, our $\mathsf{AStar}^+$-BMa algorithm outperforms the state-of-the-art algorithms by several orders of magnitude for both GED computation and GED verification. An immediate possible direction of future work is to conduct an experimental study of the existing indexing techniques for graph similarity search by adopting our much better algorithm $\mathsf{AStar}^+$-BMa for GED verification.

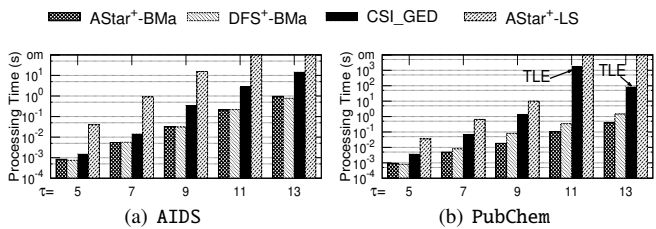

Figure 13: Against CSI_GED for GED verification (vary $\tau$)

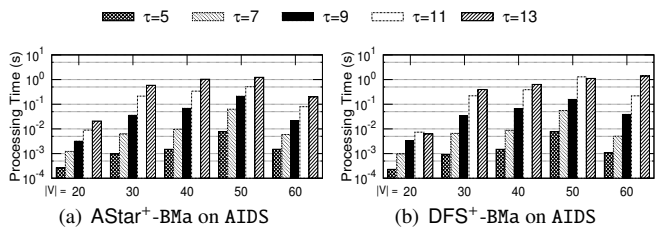

Figure 14: Scalability testing for GED verification (vary $|V|$)

[32] Gaoping Zhu, Xuemin Lin, Ke Zhu, Wenjie Zhang, and Jeffrey Xu Yu. Treespan: efficiently computing similarity all-matching. In *Proc. of SIGMOD'12*, 2012.

# A  Appendix

## A.1  A Frequency-Aware Mapping Order

The mapping order $\pi$ also has an impact on the performance of Algorithm 2, in a similar way to the impact of search order on the performance of subgraph matching algorithms (*e.g.*, see [4, 8, 12]). However, the existing techniques of computing a search order for subgraph matching cannot be applied to GED computation, since the vertex mapping in GED is unconstrained (*i.e.*, a vertex of $q$ can map to any vertex of $g$ regardless of their labels). Thus, we propose techniques to compute a mapping order for GED computation. We start with our two main intuitions.

Firstly, a connected mapping order is preferred; that is, each vertex $v$ should be connected to one of the vertices preceding $v$ in $\pi$. The intuition is that partial mappings obtained by a connected mapping order tend to have tighter lower bound costs, and the number of partial mappings enumerated by a GED computation algorithm becomes smaller if the lower bound gets tighter. Secondly, infrequent part of a graph should be mapped first. The intuition is that an infrequent subgraph of $q$ has a smaller number of similar mapping subgraphs in $g$. Consequently, most of the partial mapping generated by this infrequent subgraph have large lower bound costs, due to the large $\delta_f(q[f], g[f])$ which is a part of the lower bound cost $\delta(f)$, and thus will not be extended or generated by a GED computation algorithm. In contrast, a frequent subgraph of $q$ has a lot of similar mapping subgraphs in $g$, and thus will generate a lot of partial mappings with small lower bound costs.

Based on the above intuitions, we propose our mapping order computation algorithm, denoted by MappingOrder. To quantify the infrequency of a subgraph, we compute an infrequency weight $w(\cdot)$ for each vertex and each edge of $q$, which is one minus the label's frequency in $g$ for the corresponding vertex or edge. Thus, a subgraph is more infrequent if it has a larger total weight for vertices and edges. We use a greedy strategy to construct the mapping order $\pi$. The first vertex $v_1$ is chosen as the one with the largest total weight for the vertex and its adjacent edges. Then, we iteratively add into $\pi$ the vertex that has the largest total weight for the vertex and its adjacent edges to vertices in $\pi$.

## A.2  Edit Distance between Two Multi-sets

Given two multi-sets $S_1$ and $S_2$, the edit distance between $S_1$ and $S_2$, denoted by $\Upsilon(S_1, S_2)$, is the minimum number of edit operations that transform $S_1$ to $S_2$, where edit operations are (1) inserting an element, (2) deleting an element, and (3) replacing an element. $\Upsilon(\cdot, \cdot)$ is a metric, and $\Upsilon(\emptyset, \emptyset) = 0$ and $\Upsilon(\emptyset, S) = |S|$. In general,

$$\Upsilon(S_1, S_2) = \max\{|S_1|, |S_2|\} - |S_1 \cap S_2|,$$

where $S_1 \cap S_2$ and $S_1 \cup S_2$ denote the multi-set intersection and multi-set union, respectively. For example, if $S_1 = \{a, a, b\}$ and $S_2 = \{a, a, a\}$, the edit distance between $S_1$ and $S_2$ is 1 (*i.e.*, replace $b$ in $S_1$ with $a$); $S_1 \cap S_2 = \{a, a\}$, and $S_1 \cup S_2 = \{a, a, a, a, b\}$. It is easy to see that by using a hash structure, $\Upsilon(S_1, S_2)$ can be computed in $O(|S_1| + |S_2|)$ time.

The properties of $\Upsilon(\cdot, \cdot)$ are shown in the following two lemmas.

**Lemma A.1:** *Given four multi-sets $S_1, S_2, S'_1, S'_2$, we have $\Upsilon(S_1 \cup S'_1, S_2 \cup S'_2) \leq \Upsilon(S_1, S_2) + \Upsilon(S'_1, S'_2)$.*

**Proof:** We have

$$\Upsilon(S_1, S_2) + \Upsilon(S'_1, S'_2) - \Upsilon(S_1 \cup S'_1, S_2 \cup S'_2)$$
$$= \max\{|S_1|, |S_2|\} - |S_1 \cap S_2| + \max\{|S'_1|, |S'_2|\} - |S'_1 \cap S'_2|$$
$$\quad - \max\{|S_1 \cup S'_1|, |S_2 \cup S'_2|\} + |(S_1 \cup S'_1) \cap (S_2 \cup S'_2)|$$
$$= (\max\{|S_1|, |S_2|\} + \max\{|S'_1|, |S'_2|\} - \max\{|S_1| + |S'_1|, |S_2| + |S'_2|\})$$
$$\quad + (|(S_1 \cup S'_1) \cap (S_2 \cup S'_2)| - |S_1 \cap S_2| - |S'_1 \cap S'_2|)$$
$$\geq 0 + 0 = 0$$

Thus, the lemma holds. □

**Lemma A.2:** *Given two multi-sets $S_1, S_2$, we have $\Upsilon(S_1 \cup S_1, S_2 \cup S_2) = \Upsilon(S_1, S_2) + \Upsilon(S_1, S_2)$.*

**Proof:** This lemma can be proved in a similar way to the proof of Lemma A.1. □

The above two lemmas can also be naturally extended to the union of multiple multi-sets.

## A.3  Star Match-based Lower Bounds

**Star Match-based Lower Bound $\underline{\delta}^{\text{SM}}(\cdot, \cdot)$.** Lower bound based on the star structure is proposed in [28], where edges have no labels. We extend it to handle edge labels as follows.

*Definition A.1:* The **star** of a vertex $v$ in a graph $q$ is $S(v) = (l(v), L_E(v), L_V(v))$, where $L_V(v)$ denotes the multi-set of labels of $v$'s one-hop neighbors.

Based on the star structures $S(v)$ and $S(u)$, we define the mapping cost of mapping $v \in q \backslash f$ to $u \in g \backslash f$ as,

$$\lambda^{\text{SM}}(v, u) := \mathbb{1}_{l(v) \neq l(u)} + \tfrac{1}{2} \times \Upsilon(L_E(v), L_E(u)) + \Upsilon(L_V(v), L_V(u))$$

where $L_V(v)$ and $L_V(u)$ only consider the neighbors of $v$ and $u$ in $q \backslash f$ and $g \backslash f$, respectively. Thus, $\lambda^{\text{SM}}(v, u) = \lambda^{\text{BM}}(v, u) + \Upsilon(L_V(v), L_V(u))$. The star match-based lower bound [28] is,

$$\underline{\delta}^{\text{SM}}(q \backslash f, g \backslash f) := \frac{\min_{\sigma \in \mathcal{F}(q \backslash f, g \backslash f)} \sum_{v \in q \backslash f} \lambda^{\text{SM}}(v, \sigma(v))}{\max\{4, \Delta(q \backslash f) + 1, \Delta(g \backslash f) + 1\}}$$

where $\Delta(q \backslash f)$ and $\Delta(g \backslash f)$ denote the maximum vertex degree in $q \backslash f$ and $g \backslash f$, respectively.

**Anchor-aware Star Match-based Lower Bound.** Similarly, we revise the star structure to define the mapping cost of mapping $v \in q \backslash f$ to $u \in g \backslash f$ as,

$$\lambda^{\text{SMa}}(v, u) := \mathbb{1}_{l(v) \neq l(u)} + \tfrac{1}{2} \times \Upsilon(L_{E_I}(v), L_{E_I}(u))$$
$$\quad + \sum_{v' \in q[f]} \mathbb{1}_{l(v, v') \neq l(u, f(v'))} + \Upsilon(L_V(v), L_V(u)).$$

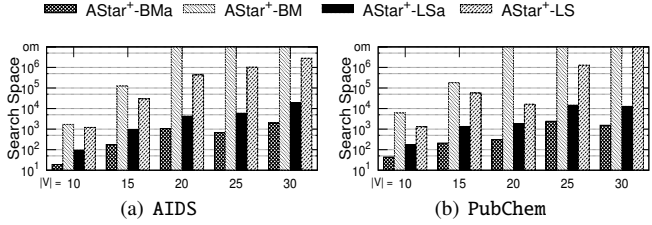
(a) AIDS  (b) PubChem
Figure 15: Evaluate anchor aware (vary |V|)

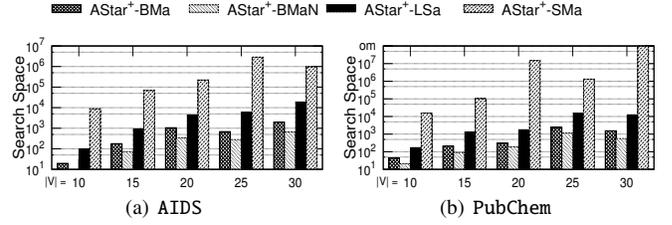
(a) AIDS  (b) PubChem
Figure 16: Evaluate lower bounds (vary |V|)

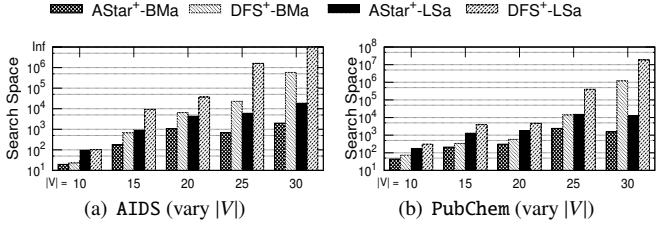
(a) AIDS (vary |V|)  (b) PubChem (vary |V|)
Figure 17: Evaluate $\text{AStar}^+$ v.s. $\text{DFS}^+$ for GED computation

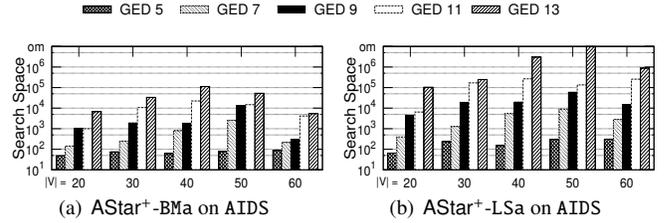
(a) $\text{AStar}^+$-BMa on AIDS  (b) $\text{AStar}^+$-LSa on AIDS
Figure 18: Scalability testing for GED computation (vary |V|)

That is, $\lambda^{\text{SMa}}(v, u) = \lambda^{\text{BMa}}(v, u) + \Upsilon(L_V(v), L_V(u))$. Then, we define the anchor-aware star match-based lower bound as,

$$\underline{\delta}^{\text{SMa}}(q\backslash f, g\backslash f) := \frac{\min_{\sigma \in \mathcal{F}(q\backslash f, g\backslash f)} \sum_{v \in q\backslash f} \lambda^{\text{SMa}}(v, \sigma(v))}{\max\{4, \Delta(q\backslash f) + 1, \Delta(g\backslash f) + 1\}}.$$

It can be easily verified that $\lambda^{\text{SMa}}(v, u) \geq \lambda^{\text{SM}}(v, u)$, and thus we have $\underline{\delta}^{\text{SMa}}(q\backslash f, g\backslash f) \geq \underline{\delta}^{\text{SM}}(q\backslash f, g\backslash f)$.

We prove the following lemma regarding $\underline{\delta}^{\text{BMa}}(\cdot, \cdot)$ and $\underline{\delta}^{\text{SMa}}(\cdot, \cdot)$.

**Lemma A.3:** *For a partial mapping $f$, we have $\underline{\delta}^{\text{BMa}}(q\backslash f, g\backslash f) \geq \underline{\delta}^{\text{SMa}}(q\backslash f, g\backslash f)$ if $\underline{\delta}^{\text{BMa}}(q\backslash f, g\backslash f) \geq |V(q\backslash f)|$.*

**Proof:** Let $d$ be $\max\{4, \Delta(q\backslash f) + 1, \Delta(g\backslash f) + 1\}$, and $\sigma$ be the mapping obtained by

$$\arg\min_{\sigma' \in \mathcal{F}(q\backslash f, g\backslash f)} \sum_{v \in q\backslash f} \lambda^{\text{BMa}}(v, \sigma'(v)).$$

Then, we have

$$d \times (\underline{\delta}^{\text{BMa}}(q\backslash f, g\backslash f) - \underline{\delta}^{\text{SMa}}(q\backslash f, g\backslash f))$$
$$= (d \times \min_{\sigma' \in \mathcal{F}(q\backslash f, g\backslash f)} \sum_{v \in q\backslash f} \lambda^{\text{BMa}}(v, \sigma'(v)))$$
$$\quad - \min_{\sigma'' \in \mathcal{F}(q\backslash f, g\backslash f)} \sum_{v \in q\backslash f} \lambda^{\text{SMa}}(v, \sigma''(v))$$
$$\geq (d \times \sum_{v \in q\backslash f} \lambda^{\text{BMa}}(v, \sigma(v))) - \sum_{v \in q\backslash f} \lambda^{\text{SMa}}(v, \sigma(v))$$
$$= \sum_{v \in q\backslash f} (d \times \lambda^{\text{BMa}}(v, \sigma(v)) - \lambda^{\text{SMa}}(v, \sigma(v)))$$

Consider each component in the last expression and let $u$ denote $\sigma(v)$. Based on the property that $\lambda^{\text{SMa}}(v, u) = \lambda^{\text{BMa}}(v, u) + \Upsilon(L_V(v), L_V(u))$, we have

$$d \times \lambda^{\text{BMa}}(v, u) - \lambda^{\text{SMa}}(v, u)$$
$$= d \times \lambda^{\text{BMa}}(v, u) - (\lambda^{\text{BMa}}(v, u) + \Upsilon(L_V(v), L_V(u)))$$
$$= (d-1) \times \lambda^{\text{BMa}}(v, u) - \Upsilon(L_V(v), L_V(u))$$
$$\geq (d-1) \times (\lambda^{\text{BMa}}(v, u) - 1)$$

where the last inequality follows from the fact that $\Upsilon(L_V(v), L_V(u)) \leq \max\{|L_V(v)|, |L_V(u)|\} \leq d - 1$.

Thus, from the above, we have

$$d \times (\underline{\delta}^{\text{BMa}}(q\backslash f, g\backslash f) - \underline{\delta}^{\text{SMa}}(q\backslash f, g\backslash f))$$
$$\geq \sum_{v \in q\backslash f} (d \times \lambda^{\text{BMa}}(v, \sigma(v)) - \lambda^{\text{SMa}}(v, \sigma(v)))$$
$$\geq (d-1) \times \sum_{v \in q\backslash f} (\lambda^{\text{BMa}}(v, \sigma(v)) - 1)$$
$$= (d-1) \times (\underline{\delta}^{\text{BMa}}(q\backslash f, g\backslash f) - |V(q\backslash f)|)$$

Therefore, if $\underline{\delta}^{\text{BMa}}(q\backslash f, g\backslash f) \geq |V(q\backslash f)|$, then $\underline{\delta}^{\text{BMa}}(q\backslash f, g\backslash f) \geq \underline{\delta}^{\text{SMa}}(q\backslash f, g\backslash f)$. □

Note that, the above lemma is conservative, while in practice, $\underline{\delta}^{\text{SMa}}(q\backslash f, g\backslash f)$ is even smaller than $\underline{\delta}^{\text{LSa}}(q\backslash f, g\backslash f)$ as verified by our experiments in Section 6. The main reason is that, as the label of a vertex $v$ is considered multiple times in the star structures of $v$'s neighbors, the mapping cost $\lambda^{\text{SMa}}(v, u)$ has to be normalized by a large factor of $\max\{4, \Delta(q\backslash f) + 1, \Delta(g\backslash f) + 1\}$.

### A.4 Additional Experimental Results

Additional experimental results regarding the search spaces of the algorithms are shown in Figure 15, Figure 16, Figure 17, Figure 18, and Figure 19.

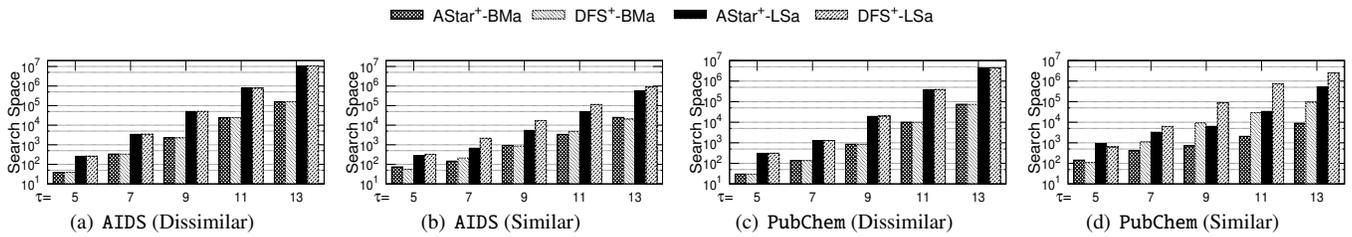

Figure 19: Evaluate AStar$^+$ v.s. DFS$^+$ for GED verification (vary $\tau$)